\documentclass[12pt, a4paper]{article}
\usepackage{amsmath,amssymb,bm,epsfig,afterpage}
\usepackage{cite}
\usepackage{here}
\usepackage{color}
\usepackage{colortbl}
\usepackage{xcolor}
\usepackage{tabularx}
\usepackage{subcaption}
\usepackage{bbm}
\usepackage{graphicx}
\usepackage{listings}
\usepackage{longtable}
\usepackage[utf8]{inputenc}

\makeatletter
    
    \@addtoreset{equation}{section}
  \makeatother

\definecolor{Orange}{cmyk}{0,0.61,0.87,0}
\definecolor{JungleGreen}{cmyk}{0.99,0,0.52,0}
\definecolor{OliveGreen}{cmyk}{0.64,0,0.95,0.40}
\definecolor{Brown}{cmyk}{0,0.81,1,0.60}
\definecolor{RoyalBlue}{cmyk}{0.71,0.53,0,0.12}
\definecolor{Gray}{cmyk}{0,0,0,0.40}
\definecolor{LightPink}{cmyk}{0.0,0.25,0,0}
\definecolor{LLightPink}{cmyk}{0.0,0.10,0,0}
\definecolor{LightBlue}{cmyk}{0.25,0,0,0}
\definecolor{LightGray}{cmyk}{0,0,0,0.2}

\newcommand{\vev}[1]{{\left\langle{#1}\right\rangle}}
\newcommand{\order}[1]{\mathcal{O}\left({#1}\right)}

\newcommand{{\Lcal}}{\mathcal{L}}

\newcommand{\PQ}{{\mathrm{PQ}}}

\newcommand{\ther}{{\mathrm{th}}}
\newcommand{\eV}{{\mathrm{eV}}}

\newcommand{\GeV}{{\mathrm{GeV}}}

\newcommand{\PeV}{{\mathrm{PeV}}}
\newcommand{\osc}{\mathrm{osc}}
\newcommand{\eff}{\mathrm{eff}}

\newcommand{\DW}{\mathrm{DW}}

\newcommand{\obs}{\mathrm{obs}}
\newcommand{\DM}{\mathrm{DM}}

\newcommand{\abs}[1]{\left|{#1}\right|}

\newcommand{\ol}[1]{\overline{#1}}
\newcommand{\wt}[1]{\widetilde{#1}}

\newcommand{\la}{\lambda}

\newcommand{\BL}{{B\mathrm{-}L}}

\newcommand{\RD}{\mathrm{RD}}
\newcommand{\MD}{\mathrm{MD}}

\newcommand{\meV}{\mathrm{meV}}

\newcommand{\LQCD}{\Lambda_{\mathrm{QCD}}}

\newcommand{\Wcal}{\mathcal{W}}

\allowdisplaybreaks[3]
\newcolumntype{Y}{&gt;{\centering\arraybackslash}X}
\usepackage[colorlinks=true, linkcolor=OliveGreen, citecolor=RoyalBlue,
urlcolor=RoyalBlue]{hyperref}

\definecolor{darkgreen}{HTML}{109930}

\begin{document}

\begin{titlepage}

\begin{flushright}
{\tt
CTPU-PTC-21-33
}
\end{flushright}

\vskip 1.35cm
\begin{center}

{\Large
{\bf
Lepto-axiogenesis in minimal SUSY KSVZ model
}
}

\vskip 1.5cm

Junichiro~Kawamura$^{a,b}$\footnote{%
\href{mailto:jkawa@ibs.re.kr}{\tt jkawa@ibs.re.kr}},
Stuart~Raby$^{c}$\footnote{%
\href{mailto:raby.1@osu.edu}{\tt raby.1@osu.edu}},
\vskip 0.8cm

{\it $^a$Center for Theoretical Physics of the Universe, Institute for Basic Science,
Daejeon, 34126, Korea}
\\[3pt]

{\it $^b$Department of Physics, Keio University, Yokohama 223-8522, Japan}
\\[3pt]

{\it $^c$Department of Physics, Ohio State University, Columbus, Ohio
 43210, USA}
\\[3pt]

\date{\today}

\vskip 1.5cm

\begin{abstract}
We study the lepto-axiogenesis scenario
in the minimal supersymmetric KSVZ axion model.
Only one Peccei-Quinn (PQ) field and vector-like fields are introduced
besides the MSSM with the type-I see-saw mechanism.
The PQ field is stabilized by the radiative correction induced
by the Yukawa couplings with the vector-like fields introduced in the KSVZ model.
We develop a way to follow the dynamics of the PQ field,
in particular we found a semi-analytical solution which describes the rotational motion
under the logarithmic potential
with including the thermalization effect via the gluon scattering
which preserves the PQ symmetry.
Based on the solution, we studied the baryon asymmetry, the effective number of neutrino,
and the dark matter density composed of the axion and the neutralino.
We found that the baryon asymmetry is successfully explained
when the mass of PQ field is $\mathcal{O}({10^6~\mathrm{GeV}})$
($\mathcal{O}({10^5~\mathrm{GeV}})$)
with the power of the PQ breaking term being $10$ ($8$).
\end{abstract}

\end{center}

 \clearpage
 \tableofcontents
 \thispagestyle{empty}
\end{titlepage}
\setcounter{footnote}{0}

\section{Introduction}

The rotational motion of a complex scalar field is considered
to be a possible source for the baryon asymmetry of the universe,
as originally considered in the Affleck-Dine (AD) baryogenesis scenario~\cite{Affleck:1984fy,Dine:1995kz}.
Flatness of a potential is key to generating a sufficient amount of asymmetry,
which is naturally explained by a flat direction in the scalar potential
of the Minimal Supersymmetric Standard Model (MSSM) in the original AD scenario.
Recently, it was proposed in Ref.~\cite{Co:2019wyp} that the asymmetry can also originate
from the rotational motion of the Peccei-Quinn (PQ) field
introduced to solve the strong CP problem~\cite{Peccei:1977hh, Peccei:1977ur}.
In this so-called axiogenesis scenario,
the PQ asymmetry is generated by the rotational motion,
and then it is readily converted to a baryon asymmetry
through sphaleron processes and perturbative interactions.
In particular, the conversion can be efficiently induced
through the Weinberg operator~\cite{Weinberg:1979sa} for neutrino masses which violates lepton number,
and this scenario is known as lepto-axiogenesis~\cite{Co:2020jtv}.

In this paper,
we study the lepto-axiogenesis scenario in the minimal supersymmetric
KSVZ axion model.
The model has only one PQ field and it has Yukawa couplings to vector-like fields,
so that the QCD anomaly is induced to solve the strong CP problem.
The PQ field is stabilized at its minimum by the potential induced radiatively
through the Yukawa couplings to the vector-like fields~\cite{Moxhay:1984am}.
Here, we consider supersymmetry (SUSY) to ensure that the scalar potential is almost flat
along the PQ field direction.
In the minimal model, the PQ field is thermalized only by gluon scattering~\cite{Bodeker:2006ij,Laine:2010cq,Mukaida:2012qn}.
This is the minimal possibility for lepto-axiogenesis with the KSVZ mechanism~\cite{Kim:1979if,Shifman:1979if}.
The model has already been  studied in Ref.~\cite{Co:2020jtv}.
In that paper the authors evaluated the dynamics of the PQ field using analytical approximations based on conservation laws,
whereas, in this paper, we examine the scenario by following the dynamics explicitly
using a numerical evaluation based on the equation of motions.
In this way, we can calculate the cosmological observables
by directly solving the evolution equations of energy and number densities.
Further, our way of calculation can be applied
for any combinations of energy densities.
We find that the radiation and PQ field energies are comparable
for substantially long times in a wide parameter space.
In such a case, the estimations based on simply radiation domination (RD)
or matter domination (MD) may not be applied.
Assuming the value of the reheating temperature after inflation,
we can calculate the gravitino density,
and hence the density of the lightest SUSY particle (LSP).
We shall also discuss the dark matter (DM) composed of the axion and the LSP.

The rest of this paper is organized as follows.
A way to follow the PQ field dynamics with rotational motion
is shown in Section~\ref{sec-PQdyn}.
In Section~\ref{sec-Cosmo},
the cosmological implications of the dynamics are discussed.
The paper is summarized in Section~\ref{sec-Summary}.
We briefly discuss the case in which the PQ field starts to rotate
during the (inflaton) MD era in Appendix~\ref{sec-HilemP}.
The importance of the thermal log potential is discussed in Appendix~\ref{sec-ThLog}.

\section{PQ field dynamics}
\label{sec-PQdyn}

\subsection{Scalar potential and initial condition}
We shall study the dynamics of the complex PQ field, $P$,
whose scalar potential is given by
\begin{align}
\label{eq-VP}
 V = m_P^2 \left(\log \frac{\abs{P}^2}{v_P^2} - 1\right)  \abs{P}^2  + m_P^2 v_P^2
   + \la^2 \frac{\abs{P}^{2(n-1)}}{M_p^{2(n-3)}}
     + \frac{A_P}{M_p^{n-3}} \left(P^n + h.c.\right) + V_H + V_\ther,
\end{align}
where the Hubble induced potential is assumed to be
\begin{align}
 V_H = - c_H H^2 \abs{P}^2,
\end{align}
with $c_H > 0$. The constant term $m_P^2 v_P^2$ is introduced
so that the potential energy in the vacuum is vanishing.
$V_\ther$ is the the thermal-log potential given by~\cite{Anisimov:2000wx},
\begin{align}
V_\mathrm{th} = a_L \alpha_s(T)^2 T^4 \log \frac{\abs{P}^2}{T^2},
\end{align}
where $a_L= 1$ is assumed in this paper.
The importance of the thermal log potential is discussed in Appendix~\ref{sec-ThLog}.
Note that the vector-like fields are heavier than the temperature throughout the dynamics
as discussed in Section~\ref{sec-dynII},
and hence the thermal mass correction is absent.
In our numerical analysis, the strong coupling constant, $\alpha_s(T)$,
is evaluated by solving the 1-loop renormalization group equation in the MSSM.

This potential is motivated by SUSY models with a superpotential,
\begin{align}
\label{eq-WP}
 W_P = y P \ol{\Psi} \Psi + \la \frac{P^n}{nM_p^{n-3}},
\end{align}
where $\Psi$, $\ol{\Psi}$ are vector-like superfields,
such that the mixed anomaly of the PQ symmetry and $SU(3)_C$ is induced~\footnote{
If the QCD anomaly is vanishing, the axion is not the QCD axion,
but an axion-like particle.
This opens up new possibilities~\cite{Co:2020xlh}, e.g. wider range of the decay constant,
but this is beyond the scope of this paper.
}.
In this work, we assume the Yukawa coupling constant $y$ to be $\order{1}$,
so that the radiative PQ breaking~\cite{Moxhay:1984am} is realized
by the logarithmic SUSY breaking mass in the first term of Eq.~\eqref{eq-VP}.
In addition, the self-coupling of the PQ field is induced
by the soft SUSY breaking effect parametrized by $A_P$~\footnote{
The factor of $1/n$, naively expected from Eq.~\eqref{eq-WP}, 
is absorbed in the normalization of $A_P$.  
}.
Since this self-coupling violates the PQ symmetry
and hence induces a mass term for the QCD axion,
the power $n$ should be sufficiently large such that
\begin{align}
\label{eq-DeltaTheta}
 \Delta \theta \sim 10^{64-10n} \times
   \left(\frac{A_P}{1~\PeV}\right)\left(\frac{v_P}{10^8~\GeV}  \right)^n
 < 10^{-10},
\end{align}
to be consistent with the measurement of neutron EDM~\cite{Baker:2006ts,Pendlebury:2015lrz,Graner:2016ses}.
Thus, $n \ge 8$ is required for the axion quality
if $v_P \gtrsim \order{10^8~\GeV}$.

We shall consider a scenario where the PQ field starts to move
when the reheating process after inflation ends,
i.e. $\dot{P} = 0$ at $T=T_i$, where $T_i$ is the reheating temperature.
The initial location of the PQ field is assumed
to be at the minimum of the potential without the SUSY breaking effects,
\begin{align}
\label{eq-Pi}
P = P_i :=
\left(\frac{c_H H^2_i M_p^{2n-6}}{(n-1)\la^2}\right)^{\frac{1}{2n-4}}
 e^{i \theta_i},
\end{align}
where $H_i$ is the Hubble constant at the initial time.
The initial angle $\theta_i$ is not fixed,
since the potential does not depend on the angle unless the A-term is sizable.
The initial value is estimated as~\footnote{
For the numerical values in this section,
we take $g_*(T_i) = g_*^\mathrm{MSSM} = 228.75$, $c_H = \la = 1$  and $n=10$,
although we keep this in the analytical expressions.
We assume $g_*(T)=g_*^\mathrm{MSSM}$ throughout the paper
except discussions about $\Delta N_\eff$  and the DM in Section~\ref{sec-DM}.
}
\begin{align}
\label{eq-Si}
\abs{P_i}\sim \left( \frac{\pi^2 g_*(T_i) c_H}{90(n-1)\la^2}
       T_i^4 M_p^{2n-8} \right)^{\frac{1}{2n-4}}
     \sim 6.5 \times 10^{16}~\GeV \times
           \left(\frac{T_i}{10^{12}~\GeV}\right)^{1/4}.
\end{align}
Here, we assume that the radiation energy dominates the universe.
The initial value increases slightly as $T_i$ increases and/or the power $n$ is smaller.
If $H_i < m_P$,
the PQ field starts to rotate around the minimum during the (inflaton) MD era.
We need to specify how the reheating proceeds to numerically follow the dynamics
which is beyond the scope of this paper.
Hence, we shall focus on the case of $H_i > m_P$.
The case of $H_i < m_P$ is briefly discussed in Appendix~\ref{sec-HilemP}.

\subsection{PQ dynamics when $H \gtrsim m_P$}
\label{sec-dynI}

The PQ field is kicked by the A-term in the early stage when the amplitude is large.
The equation of motion is given by
\begin{align}
&\  \ddot{P} + 3 H \dot{P} + \frac{\partial V}{\partial P^*} = 0,   \\
&\ \frac{\partial V}{\partial P^*}
= \left(m_P^2 \log \frac{\abs{P}^2}{v_P^2}- c_H H^2 + (n-1) \la^2
   \frac{\abs{P}^{2(n-2)}}{M_p^{2n-6}}
  + a_L \alpha_s^2 \frac{T^4}{\abs{P}^2}
\right) P  + n \frac{A_P}{M_p^{n-3}} P^{* (n-1)}.
\end{align}
It is convenient to introduce a dimensionless variable
\begin{align}
 u := \log \frac{a}{a_i},
\end{align}
where $a$ is the scale factor of the universe and $a_i$ is its initial value
at $t=t_i$.

We parametrize $P$ as
\begin{align}
 P =: \frac{S}{\sqrt{2}} e^{i\theta}
   =: \abs{P_i} \chi e^{i \theta},
\end{align}
where $S$, $\chi$, $\theta$ are real functions of $u$.
Here, $S$ and $\theta$ are respectively
the radial and angular directions of the PQ field $P$.
The evolution equations for $\chi$ and $\theta$ are given by
\begin{align}
\label{eq-eqchith}
&\  \chi^{\prime\prime} + \left(3 + \frac{H^\prime}{H} \right) \chi^\prime
  - \left(c_H + \theta^{\prime 2}\right)\ \chi  \\ \notag
 &\ + \frac{\chi}{H^{2}} \left[
  m_P^2 \left(\log \frac{\abs{P_i}^2\chi^2}{v_P^2} \right)
              + a_L \alpha_s^2 \frac{T^4}{\abs{P_i}^2 \chi^2}
              + c_H H_i^2 \chi^{2n-4} + H_i^2 a_P \chi^{n-2} \cos n\theta
  \right]  = 0, \\
&\  \theta^{\prime\prime}
 + \left(3+\frac{H^\prime}{H} \right) \theta^\prime
 + 2 \theta^{\prime} \frac{\chi^\prime}{\chi}
      - \frac{H_i^2}{H^2} a_P \chi^{n-2} \sin n\theta  = 0,
\end{align}
where
\begin{align}
\label{eq-cs}
a_P := \frac{A_P}{H_i} \sqrt{\frac{c_H n^2}{(n-1) \la^2 }}.
\end{align}
Here, $\chi^\prime$ etc. denotes the derivative with respect to $u$.
The Hubble parameter $H$ is given by
\begin{align}
 3 M_p^2 H^2 = \rho_R + \rho_P,
\quad
\rho_R = \frac{\pi^2 g_*}{30} T_i^4 e^{-4u},
\quad
\rho_P = \abs{\dot{P}}^2 + V.
\end{align}
It is convenient to introduce the notation,
\begin{align}
 y_0 = \chi,\quad
 y_1 = \frac{H}{H_i} e^{3u} \chi^\prime, \quad
 y_2 = \theta, \quad
 y_3 = \frac{H}{H_i} e^{3u} \theta^\prime,
\end{align}
so that the evolution equations for $y_a$ ($a=0,1,2,3$),
\begin{align}
\label{eq-deqys}
 y_0^\prime =&\ \frac{H_i}{H} e^{-3u} y_1, \quad
 y_2^\prime = \frac{H_i}{H} e^{-3u} y_3,  \\   \notag
 y_1^\prime =&\
  \frac{H_i}{H}y_0 \Biggl[ e^{-3u} y_3^2 \Biggr.  \\ \notag
&\ \left.  \hspace{-0.3cm}+ e^{3u}
\left\{
  c_H \frac{H^2}{H_i^2}
 -\left( \frac{m_P^2}{H^2_i} \log \frac{\abs{P_i}^2 y_0^2}{v_P^2} + c_H y_0^{2n-4} +
         a_P y_0^{n-2}\cos n y_2  +
   a_L \alpha_s^2 \frac{T_i^4 e^{-4u}}{H_i^2 \abs{P_i}^2 y_0^2}
\right)
 \right\}
 \right], \\ \notag
 y_3^\prime =&\
  \frac{H_i}{H} \left( -2 e^{-3u}  \frac{y_1y_3}{y_0} + e^{3u} a_P y_0^{n-2}
  \sin ny_2\right),
\end{align}
are independent of $H^\prime$.
With this parametrization,
\begin{align}
 \abs{\dot{P}}^2 = H_i^2 \abs{P_i}^2 e^{-6u} \left( y_1^2  + y_0^2 y_3^2\right),
\quad
  V = V(\abs{P_i} y_0 e^{iy_2}).
\end{align}
We numerically solve Eq.~\eqref{eq-deqys}
up to $H^2 = 10^{-4}\times m_P^2$,
where the Hubble constant becomes negligible compared to the mass,
and $P$ oscillates around its minimum very fast per unit Hubble time.
We define the value of $u$ at this time as $u_1$,
 i.e. $H(u_1) := 0.01\times m_P$.

\begin{figure}[t]
\centering
\begin{minipage}[c]{0.49\hsize}
\centering
\includegraphics[height=65mm]{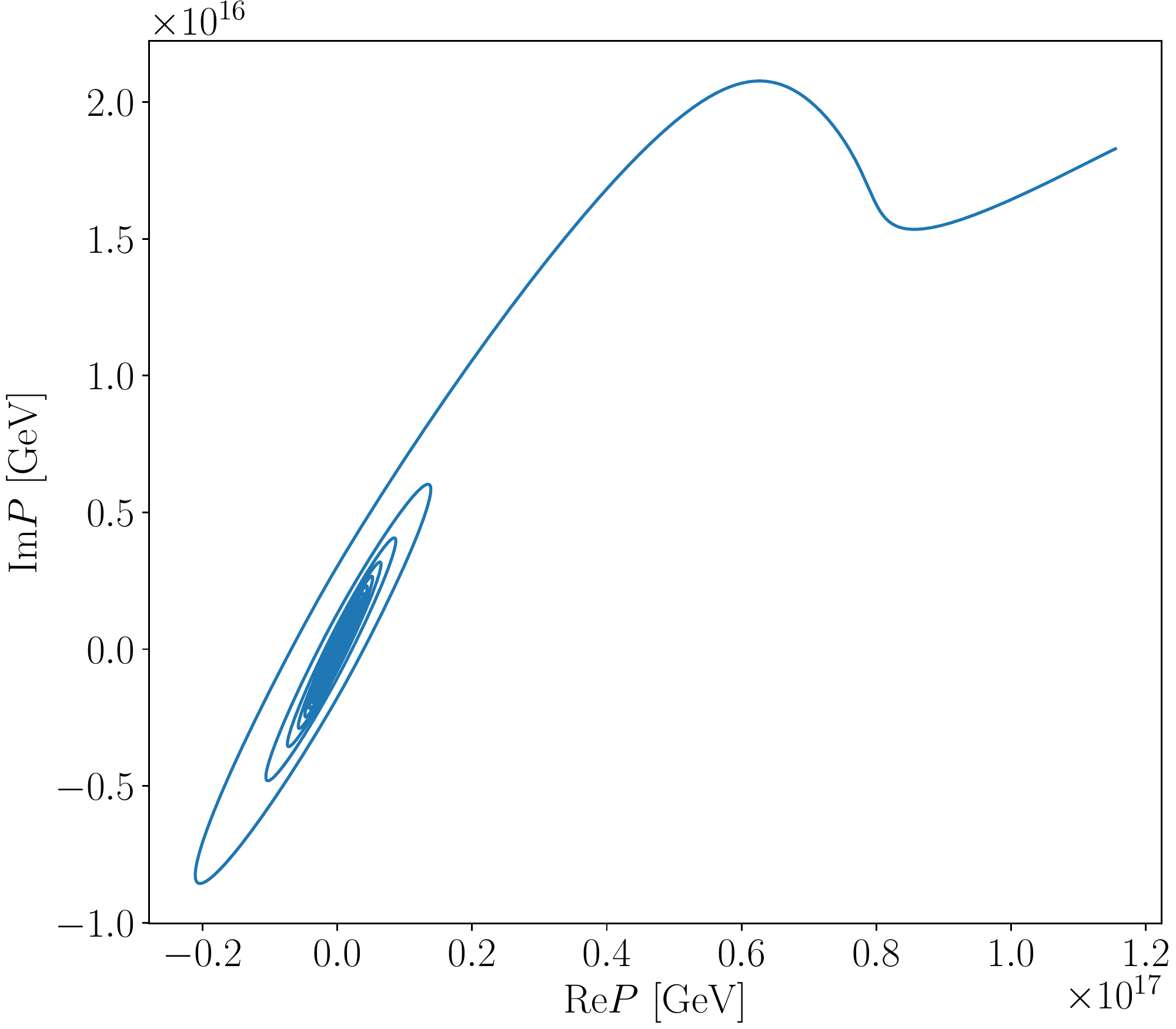}
\end{minipage}
\begin{minipage}[c]{0.49\hsize}
\centering
\includegraphics[height=65mm]{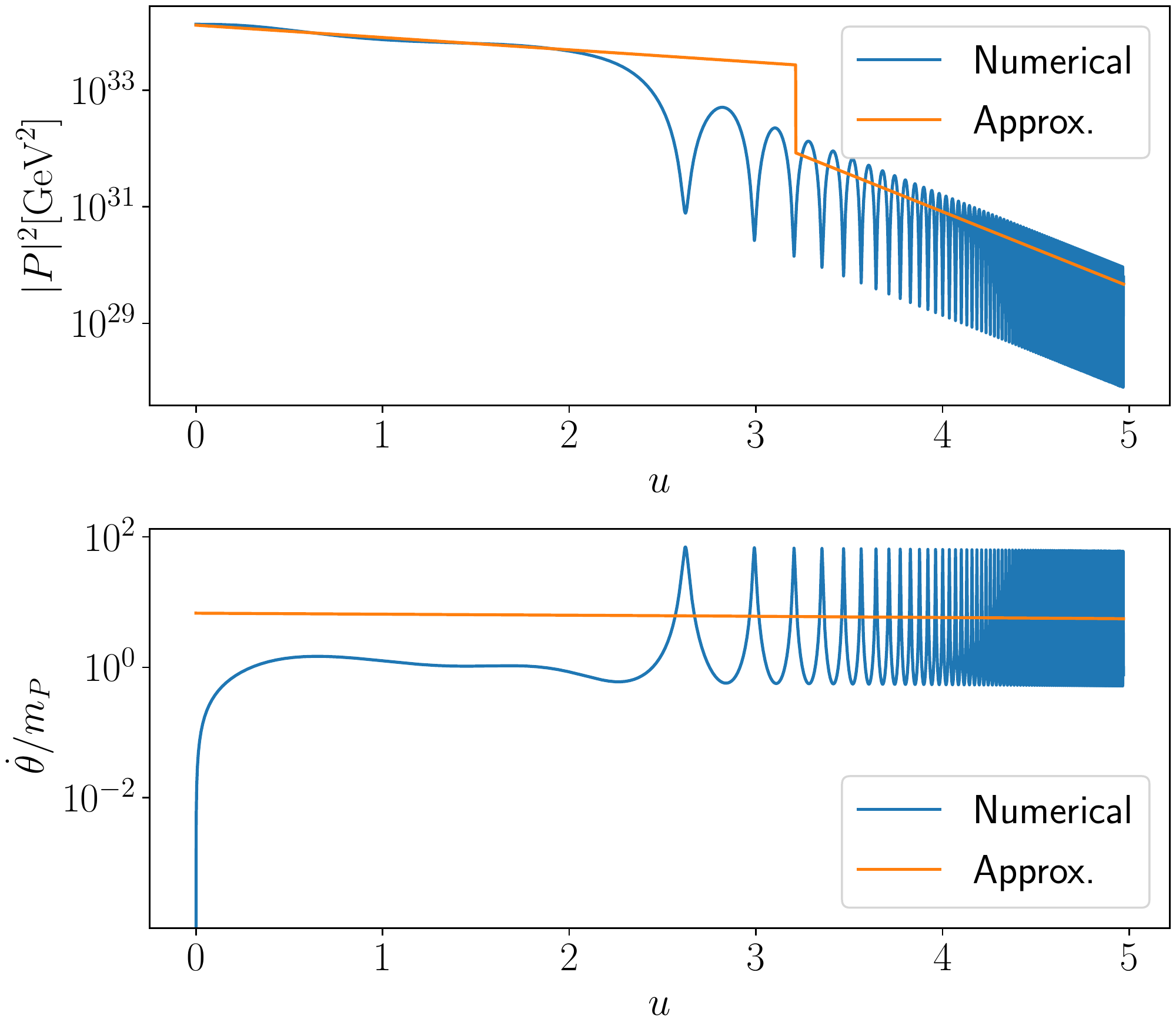}
\end{minipage}
\caption{\label{fig-dynamicsI}
The dynamics of the PQ field when the parameters are given by
$\theta_i=\pi/20$, $T_i = 10^{13}$ GeV, $v_P = 10^8$ GeV and $m_P = 10^6$ GeV.
The left panel is the plot on $(\mathrm{Re} P,~\mathrm{Im} P)$ plane.
The upper (lower) right panel shows $\abs{P}^2$ ($\dot{\theta}/m_P$).
The blue lines are the numerical solution,
and the yellow lines are the approximate solution.
}
\end{figure}

A result of numerical evaluation is shown in Fig.~\ref{fig-dynamicsI}.
The blue curves are the results of numerical evaluation,
and the yellow lines in the right panels show the approximate solution.
Before the rotation around the minimum starts at $u=u_\osc$ where $H=m_P$,
the PQ field locates around the minimum of the potential,~\footnote{Note, $u_1 > u_\osc$.}
\begin{align}
\label{eq-PappBef}
 P \sim
\left(\frac{c_H H^2 M_p^{2n-6}}{(n-1)\la^2}\right)^{\frac{1}{2n-4}} e^{i \theta}.
\end{align}
The value $\abs{P}^2$ at $u<u_\osc$ is given by this equation.
Its value at $u>u_\osc$ and $\dot{\theta}$ are extrapolated
from the approximate solution at $u > u_1$ derived in the next section.
The approximate solution roughly agrees with the numerical solutions~\footnote{
Note, however, that the solution at $u > u_1$ uses
the result of the numerical evaluation at $u < u_1$,
so this is not a fully analytical result.
}.

We see that the PQ field is kicked around $P \sim (8+1.5 i)\times 10^{16}~\GeV$,
and then starts to rotate around the minimum.
The PQ number density, $n_\PQ := i(\dot{P}^* P - P^* \dot{P})$, is generated by this kick.
The evolution equation of $n_\PQ$ is given by
\begin{align}
 \dot{n}_\PQ + 3 H n_\PQ = 2 n \frac{A_P}{M_p^{n-3}} \abs{P}^n \sin n\theta.
\end{align}
Solving this, the generated PQ number density is given by
\begin{align}
\label{eq-IntnPQ}
 n_\PQ = e^{-3u} \int^u_0 du^\prime \Delta n_\PQ(u^\prime),
\quad
\Delta n_\PQ(u)  :=  2 n \frac{A_P}{M_p^{n-3}}
               e^{3u} \frac{\abs{P}^n}{H} \sin n\theta.
\end{align}
The scaling of the integrated function in Eq.~\eqref{eq-IntnPQ}, $\Delta n_\PQ$
are as follows.
Before the oscillation starts, $u < u_\osc$,
$\Delta n_\PQ \propto e^{3u} H^{2/(n-2)}$,
so the exponent is positive for $n\gtrsim 8$, whatever the dominant energy.
While after the rotation starts, $u > u_\osc$,
$\Delta n_\PQ \propto e^{3u} \abs{P}^n/H \propto e^{3(2-n)u/2}/H$
has the negative exponent with assuming $|P|\propto e^ {-3/2 u}$~\footnote{
We will derive this in the next section.
}.
The PQ charge yield, $Y_\PQ := n_\PQ/s$, where $s$ is the entropy density,
generated around $u\sim u_\osc$ is estimated as
\begin{align}
\label{eq-Yosc-app}
 Y_\osc :=Y_\PQ (u_\osc) \sim &\
           \frac{n^2}{6n-20} \left(\frac{c_H}{(n-1)\la^2}\right)^{\frac{n}{2n-4}}
           \left(\frac{90}{\pi^2g_*}\right)^{\frac{1}{4}}
           A_P \left( \frac{M_p^{n-6}}{m_P^{3n-10}}\right)^{\frac{1}{2n-4}}
           \sin \left(n\theta_\osc\right),
\notag             \\
       \sim&\ 350 \times \left(\frac{1~\PeV}{m_P}\right)^{5/4}
                          \left(\frac{A_P}{1~\PeV}\right)
                         \; \sin \left(10 ~\theta_\osc\right),
\end{align}
where we used Eq.~\eqref{eq-PappBef} to estimate $\abs{P}$ at $u=u_\osc$
and $n=10$ is chosen in the second line.
Here, radiation energy domination is assumed.
The estimation Eq.~\eqref{eq-Yosc-app} tends to overestimate its value,
because the value of $\abs{P}$ is typically smaller than the estimation Eq.~\eqref{eq-PappBef},
as we see from Fig.~\ref{fig-dynamicsI}.
Since $Y_\PQ$ depends on $\abs{P}^n$,
even small deviations makes large differences due to the large power $n$.
Thus, we need to numerically evaluate $Y_\PQ$ by solving the evolution equation.
In fact, we will see that $Y_\PQ \sim \order{10}$ is obtained
from the numerical evaluation for sufficiently large $T_i$.

The numerical evaluation becomes ineffective for later times
due to the extremely rapid oscillation per Hubble time.
Thus we invoke a semi-analytical solution to evaluate the dynamics for $u > u_1$.

\subsection{PQ dynamics at $H \ll m_P$}
\label{sec-dynII}

At $u > u_1$, $H\ll m_P$, the higher-dimensional terms are negligible.
The evolution equations of $S$ and $\theta$ are given by
\begin{align}
%{eq-deqS} {eq-deqq}
\label{eq-deqS}
& \ddot{S} -\dot{\theta}^2 S + 3 H\dot{S} + m_P^2 S \log \frac{S^2}{2v_P^2}
   + a_L \alpha_s^2 \frac{2T^4}{S}
  =- \Gamma \dot{S},  \\
& \ddot{\theta} S + 2\dot{\theta} \dot{S}+ 3H\dot{\theta} S = 0.
\label{eq-deqq}
\end{align}
Here, we phenomenologically introduce the decay term
on the right-hand side of Eq.~\eqref{eq-deqS},
so that the radial direction loses its energy via thermalization.
As discussed later, we shall consider that the PQ field is dominantly thermalized
via gluon scattering and the thermalization rate is proportional to $T^3/S^{2}$~\cite{Bodeker:2006ij,Laine:2010cq,Mukaida:2012qn}.
We introduce the thermalization term only to the radial direction
since this thermalization process preserves the PQ symmetry.
From Eq.~\eqref{eq-deqq}, the PQ number density
$n_\PQ = i(\dot{P}^* P -P^* \dot{P}) = \dot{\theta} S^2$
is conserved up to the Hubble expansion, i.e. $\dot{n}_\PQ + 3H n_\PQ = 0$,
and a non-zero term on the right-side would violate the PQ symmetry.

Equations~\eqref{eq-deqS} and~\eqref{eq-deqq} are equivalent to
\begin{align}
\label{eq-eqpsiwg}
 \psi^{\prime\prime} + \left(3+\frac{H^\prime}{H}\right) \psi^\prime
 + \frac{m_P^2}{H^2}
 \left(\log{\abs{\psi}^2}+ \frac{\zeta(T)}{\abs{\psi}^2} \right)\psi
     = -\frac{\Gamma}{H} \frac{\abs{\psi}^\prime}{\abs{\psi}} \psi,
\end{align}
where $\psi := P/v_P$ and
\begin{align}
 \zeta(T) := a_L \alpha_s(T)^2 \frac{T^4}{m_P^2 v_P^2}.
\end{align}
$\psi$ and $\psi^\prime$ are related to $\chi$, $\theta$ as
\begin{align}
\label{eq-init1}
 \psi_1 =&\ \frac{\abs{P_i}}{v_{P}} \chi_1 e^{i\theta_1 },
\quad
 \psi^\prime_1 = \frac{\abs{P_i}}{v_{P}}
       \left(\chi^\prime_1  + i \theta^\prime \chi_1 \right)  e^{i\theta_1},
\end{align}
where $F_1 := F(u_1)$ for $F = \psi, \chi, \theta$ and their derivatives.

We can derive the approximate solution for the evolution equation Eq.~\eqref{eq-eqpsiwg}.
We consider the ansatz
\begin{align}
\label{eq-solpsi}
 \psi = e^{\Omega_+ + i B_+} + e^{\Omega_- -i B_-},
\end{align}
where $\Omega_\pm$ and $B_\pm$ are real positive functions,
and then we introduce real functions $f_\pm$,
\begin{align}
 B_\pm^\prime =: \frac{m_P}{H} f_\pm.
\end{align}
We assume that $f_\pm$ and $\Omega_\pm$ do not grow as fast as $H^{-1}$.
Neglecting the terms not enhanced by $m_P/H$,
\begin{align}
\label{eq-e2uk}
&\  \left[\left(  \log\abs{\psi}^2+\frac{\zeta(T)}{\abs{\psi}^2}-f_+^2 \right)\frac{m_P^2}{H^2}
+ i \left(  f_+^\prime + 3 f_+ + 2\Omega_+^\prime f_+  \right) \frac{m_P}{H}
\right] e^{\Omega_+ + iB_+} \\ \notag
&\ + \left[
\left( \log\abs{\psi}^2+\frac{\zeta(T)}{\abs{\psi}^2} - f_-^2 \right)  \frac{m_P^2}{H^2}
-i \left(  f_-^\prime + 3 f_- + 2\Omega^\prime_- f_-  \right) \frac{m_P}{H}
\right]  e^{\Omega_- - iB_-}  \\ \notag
&\  =
-i \frac{m_P\Gamma}{2H^2 \abs{\psi}^2} (f_+ +f_-)
\\ \notag
&\hspace{3cm} \times
\left\{
e^{2\Omega_-} \left(1-e^{-2i(B_+ +B_-)}\right) e^{\Omega_+ + iB_+}
 - e^{2\Omega_+} \left(1-e^{2i(B_+ +B_-)}\right) e^{\Omega_- - iB_-}
 \right\}.
\end{align}
Taking the real and imaginary parts of the coefficients of $e^{\Omega_\pm \pm iB_\pm}$,
\begin{align}
\label{eq-deqOmega}
& \log\abs{\psi}^2+\frac{\zeta(T)}{\abs{\psi}^2} -f_\pm^2=
         \frac{\Gamma}{2m_P} \frac{f_+ + f_- }{\abs{\psi}^2}
         e^{2\Omega_\mp} \sin2\phi, \\
& f_\pm^\prime + 3f_\pm + 2\Omega^\prime_\pm f_\pm
   = - \frac{\Gamma}{2H} \frac{f_+ + f_-}{\abs{\psi}^2}
                   e^{2\Omega_\mp} \left(1-\cos2\phi\right), %\\
\label{eq-deqfpm} %
\end{align}
where $\phi := B_+ + B_-$.
We define $\Omega_\pm =: \Omega \pm \Delta/2$,
and a real function $\gamma$ whose first derivative is given by
\begin{align}
\label{eq-gammap}
 \frac{\Gamma}{H} =: \frac{2 e^{2\Omega_+} \gamma^\prime}{\abs{\psi}^2}
                  = \frac{\gamma^\prime e^{\Delta}}{\cosh \Delta + \cos \phi}.
\end{align}
We factorize the oscillating part (and $e^\Delta$) of the thermalization rate
from the other parts.
Since the oscillating motion  is very fast, $\sim m_P/H$,
we replace the $\phi$ dependent parts by their averaged values, e.g.
\begin{align}
 \vev{\log{\abs{\psi}^2}}_\phi :=&\
\frac{1}{2\pi}\int^{\pi}_{-\pi} d\phi  \log \abs{\psi}^2
      = 2 \Omega + \Delta,
\end{align}
and similarly,
\begin{align}
% \vev{\log{2(\cosh\Delta + \cos\phi)}}_\phi =&\  \Delta,  \\
 \vev{\frac{\sin 2\phi}{\left(\cosh\Delta + \cos\phi\right)^2}}_\phi = 0,
~
 \vev{\frac{1-\cos2\phi}{\left(\cosh\Delta + \cos\phi\right)^2}}_\phi
   =2 \left(\coth\Delta - 1 \right),
~
\vev{\frac{1}{\abs{\psi}^2}} = \frac{e^{-2\Omega}}{2\sinh\Delta}. 
\end{align}
Then Eqs.~\eqref{eq-deqOmega} and~\eqref{eq-deqfpm}
are arranged to
\begin{align}
\label{eq-dfeqs}
\Delta^\prime = \gamma^\prime,
\quad
f^\prime + 3 f + 2 \Omega^\prime f = - f \left(\log (\sinh\Delta) \right)^\prime,
\quad
 f^2 = 2\Omega + \Delta + \frac{\zeta(T) e^{-2\Omega}}{2\sinh\Delta},
\end{align}
where $f := f_+ = f_-$.
As discussed in Appendix~\ref{sec-ThLog},
the thermal-log term in the last equation is negligible or sub-dominant
in most of the parameter space,
and thus we omit this term hereafter~\footnote{
The thermal-log term is at most 3\% of the mass term for $n=10$.
For $n=8$,
the thermal-log potential is smaller than 50\% of the mass term
when $T_i \lesssim 10^{12}~\GeV$,
but can be larger than the mass term for higher $T_i$, as shown in Fig.~\ref{fig-RT},
We would need to solve Eq.~\eqref{eq-dfeqs} numerically
to account for the thermal-log effect at $u>u_1$,
but this is beyond the scope of this paper.
}.
The solutions for $f$ and $\Omega$ are given by
\begin{align}
\label{eq-Asol}
 f = \sqrt{\frac{w}{2}}, \quad
\Omega =  \frac{w}{4} - \frac{\Delta}{2},
\quad
 w:=\Wcal \left(
 C_w^2 \left(\frac{1+\coth\Delta}{2}\right)^2 e^{-6(u-u_1)}\right),
\end{align}
where $\Wcal(z)$ is the Lambert function, which satisfies $\Wcal(z)e^{\Wcal(z)} = z$.
The approximate behavior of the Lambert function is given by
\begin{align}
 \Wcal(z) \sim
\begin{cases}
 z - z^2 + \order{z^3} & z < e^{-1} \\
\log z - \log\log z + \order{\dfrac{\log \log z}{\log z}} & z \gtrsim 3
\end{cases}.
\end{align}
Since the Lambert function does not grow exponentially,
this solution meets the assumption.
The function $\Delta$ is a solution for the differential equation,
\begin{align}
\label{eq-deqDelta}
 \Delta^\prime = \frac{\abs{\psi}^2}{2e^{w/2}} \frac{\Gamma}{H},
\quad
  C_\Delta:= \Delta(u_1).
\end{align}
Note that the right-hand side is a function of $\Delta$.
We solve this equation numerically together
with the evolution equations of the energy densities discussed in the next section.
Qualitatively, $\Delta \simeq C_\Delta$ for $\Gamma \ll H$,
while $\Delta \to \infty$ for $\Gamma \gtrsim H$.

Altogether, the approximate solution is given by
\begin{align}
\label{eq-semianal}
 \psi = e^{w/4} \left( e^{iB_+}
+ e^{-\Delta -iB_-} \right),
\end{align}
with
\begin{align}
 B_\pm = C_\pm +  \int^u_{u_1} du^\prime \frac{m_P}{H} \sqrt{\frac{w}{2}}.
\end{align}
Here, $C_\pm$, $C_\Delta$ and $C_w$ are arbitrary real constants obtained from the integrations;
to be determined by the initial condition at $u=u_1$.
With this solution, the PQ charge is given by~\footnote{
$\order{\Gamma v_P^2}$ term is
proportional to $\Gamma \abs{\psi}^2 \sin \phi$,
so it is vanishing after averaging and negligible.
}
\begin{align}
\label{eq-nPQanal}
 n_\PQ = \sqrt{2} m_P v_P^2 C_w  e^{-3(u-u_1)}
         + \order{\Gamma v_P^2},
\end{align}
so the constant $C_w$ is determined from the PQ charge,
\begin{align}
\label{eq-Cw}
 &\ C_w = \frac{n_{\PQ}(u_1)}{\sqrt{2} v_P^2 m_P}
= \frac{H_1}{\sqrt{2}m_P} \times {i(\psi^{\prime *}_1 \psi_1-\psi^\prime_1\psi^*_1)}.
\end{align}
The other constants $C_\Delta$ and $C_\pm$ are determined from
\begin{align}
C_\Delta
=\log \abs{\frac{\psi^\prime_1 +\left(\dfrac{3w_1}{2(1+w_1)}
                          + i \dfrac{m_P}{H_1}\sqrt{\dfrac{w_1}{2}}\right) \psi_1}
                         {\psi^\prime_1 +\left(\dfrac{3w_1}{2(1+w_1)}
                          - i \dfrac{m_P}{H_1}\sqrt{\dfrac{w_1}{2}}\right) \psi_1}
},
\end{align}
and
\begin{align}
 C_\pm = \pm \mathrm{Arg}\left( \mp i
\left[\psi_1^\prime + \left(\frac{3w_1}{2(1+w_1)}
  \pm i \frac{m_P}{H_1}\sqrt{\frac{w_1}{2}}\right) \psi_1 \right]
\right),
\end{align}
where $\Gamma/H$ is neglected at $u=u_1$.
Here we assume that the PQ number density is positive
as is necessary to produce the correct baryon asymmetry.

\begin{figure}[t]
\centering
\includegraphics[height=70mm]{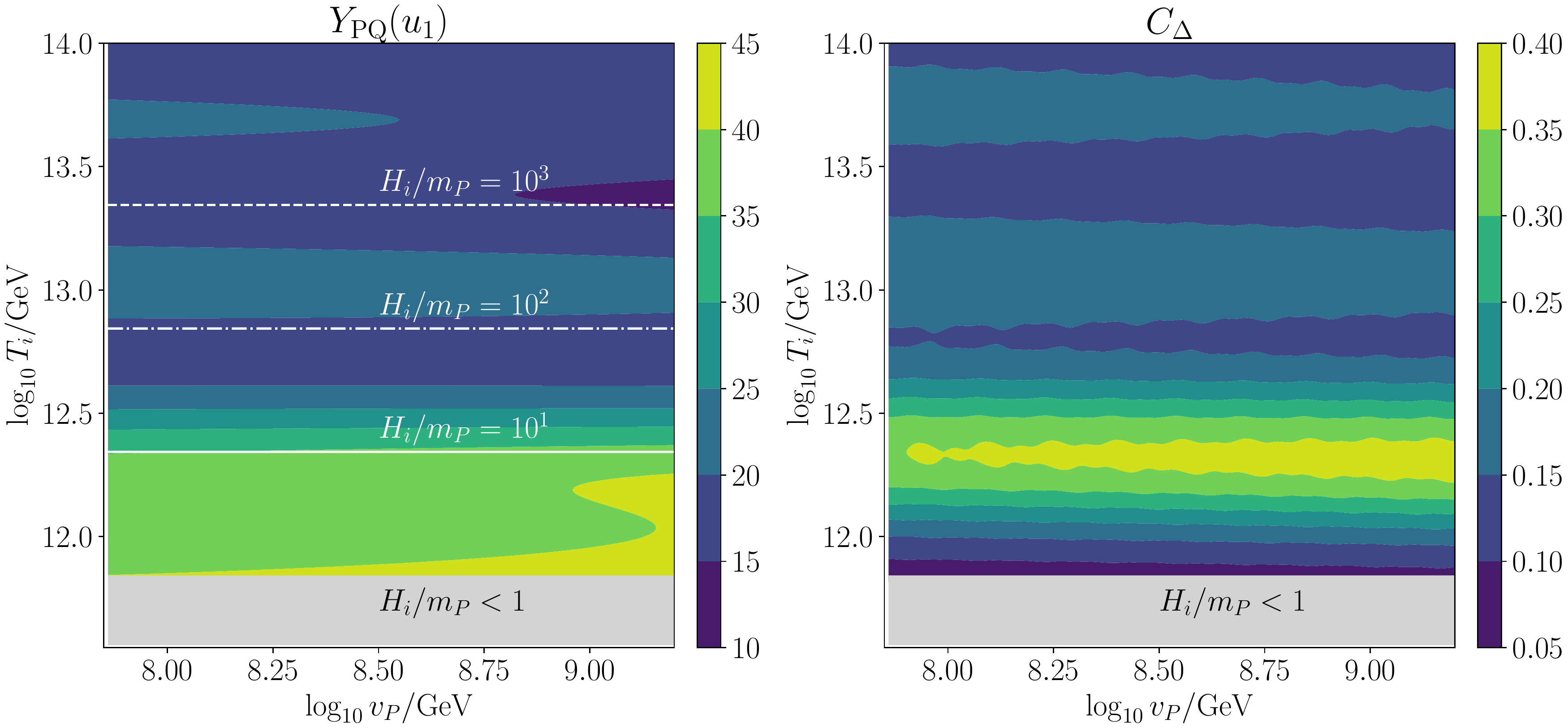}
\caption{\label{fig-Y1Cd}
Values of $Y_\PQ(u_1)$ (left) and $C_\Delta$ (right),
where $n=10$, $\theta_i=\pi/20$ and $m_P = 10^6$ GeV.
$H_i/m_P$ is shown by the white lines in the left panel.
}
\end{figure}

The values of $Y_\PQ(u_1)$ and $C_\Delta$ on $(v_P, T_i)$ plane
are shown in Fig.~\ref{fig-Y1Cd}.
In this figure, $n=10$, $\theta_i=\pi/20$ and $m_P = 10^6$ GeV.
The value of $C_w$ is determined from $Y_\PQ(u_1)$ through Eq.~\eqref{eq-Cw}.
The white lines in the left panel shows $H/m_P$.
We see that $Y_\PQ(u_1) \sim \order{10}$ at $T_i \gtrsim 10^{12}~\GeV$
where $H_i \gtrsim m_P$.
The value of $C_\Delta$ is $\order{0.1}$ at $T_i \gtrsim 10^{12}~\GeV$
and the largest value is about $0.4$.

The form of the solution Eq.~\eqref{eq-solpsi}
can be understand as a linear combination of the positively rotating mode $e^{iB_+}$
and the negatively rotating mode $e^{-iB_-}$.
While $\Gamma \ll H$ and $\Delta\lesssim\order{1}$,
both positive and negative rotational motions exist
and hence the motion is elliptic.
The minimum radius per rotation is given by~\footnote{
If the masses of vector-like fields become lighter than the temperature,
i.e. $m_\mathrm{VL} = y S_\mathrm{min}/\sqrt{2} < T$,
the thermal effects from the vector-like particles
should be taken into account~\cite{Moroi:2012vu,Moroi:2013tea,Moroi:2014mqa}.
However, we did not find any point in the parameter space where
this happens during the dynamics.
}
\begin{align}
 \frac{S_\mathrm{min}}{\sqrt{2}}= v_P e^{w/4} \left(1-e^{-\Delta}\right).
\end{align}
Note, the motion is circular for larger $\Delta$,
whereas it is more elliptic for smaller $\Delta$.
In particular, the PQ field is simply oscillating along one direction
if $\Delta \ll 1$.
This can happen if the kick effect via the A-term is not sufficiently large
due to the too small initial amplitude.
However, the produced baryon asymmetry may be too small,
so we are not interested in such a case.
After the thermalization, $\Gamma \gtrsim H$ and $\Delta \to \infty$,
the negative rotation ceases and hence the motion becomes circular.

We shall calculate the averaged values of the various quantities based on the solution.
The averaged values of $\abs{P}^2$ and $|{\dot{P}}|^2$ are given by
\begin{align}
\label{eq-PsqAvg}
 \vev{\abs{P}^2} = v_P^2 e^{w/2} \left(1+e^{-2\Delta}\right),
\quad
\rho_{\dot{P}} := \vev{\abs{\dot{P}}^2} = v_P^2 m_P^2 w e^{w/2}
                \frac{1+e^{-2\Delta}}{2}
                 + \order{v_P^2 H^2, v_P^2 \Gamma^2}.
\end{align}
Hereafter, we omit the negligible contributions
of $\order{v_P^2H^2, v_P^2 \Gamma^2}$ appeared in the derivatives.
The potential energy is given by
\begin{align}
 \rho_V :=  &\ \vev{V} =  m_P^2 v_P^2 \left[e^{w/2} \left(1+e^{-2\Delta}\right)
  \left(\frac{w}{2} - \tanh\Delta\right)      + 1 \right], % \\
\end{align}
where
\begin{align}
 \vev{\cos\phi~\log\left(\cosh\Delta+\cos\phi\right)}_\phi = e^{-\Delta}
\end{align}
is used.
Thus, the total energy of the PQ field is given by
\begin{align}
\label{eq-rhoP}
 \rho_P := \rho_{\dot{P}} + \rho_V
 = m_P^2 v_P^2 \left[e^{w/2} \left(1+e^{-2\Delta}\right)
  \left( w - \tanh\Delta \right)      + 1  \right].
\end{align}
The PQ field energy is vanishing by the red-shift, $w\to 0$
and the thermalization $\Delta \to \infty$,
since $\rho_P \to e^{-2\Delta}$ when $w \to 0$.
The energies of the radial and rotational motions are respectively given by
\begin{align}
\label{eq-rhoSdot}
\rho_{\dot{S}} :=&\ \vev{\frac{1}{2}\dot{S}^2} = v_P^2 m_P^2 w e^{w/2} e^{-2\Delta},    \\
\rho_{\dot{\theta}}:=&\ \vev{\frac{1}{2}\dot{\theta}^2S^2}
                   = \vev{\frac{n_\PQ^2}{2S^2}}
                   = v_P^2 m_P^2 w e^{w/2} \frac{1-e^{-2\Delta}}{2},
\end{align}
so we see that $\rho_{\dot{P}} = \rho_{\dot{S}} + \rho_{\dot{\theta}}$.
Qualitatively, the kinetic energy of radial motion is lost by the thermalization,
while the rotational energy is slightly increased due to the shrinking of the radius.
In particular, most of the kinetic energy is lost by the thermalization
when $C_\Delta \ll 1$ and the motion is highly elliptic.
The ratio of the energy after to before the thermalization is given by
\begin{align}
 \frac{\left.\rho_P\right|_{\Delta \to \infty}}{\left.\rho_P\right|_{\Delta = C_\Delta}}
  = \sqrt{\frac{w_{\Delta \to \infty}}{w_{\Delta = C_\Delta}}} \coth C_\Delta,
\end{align}
so only a fraction of the energy is lost if $C_\Delta$ is not so small,
while most of the energy is lost if $C_\Delta \ll 1$.
Finally, the angular velocity is given by
\begin{align}
 \vev{\dot{\theta}} = \vev{\frac{n_\PQ}{2v_P^2 \abs{\psi}^2}}
     = \frac{n_\PQ}{4v_P^2} e^{-w/2 + \Delta}  \vev{\frac{1}{\cosh\Delta+\cos\phi}}_\phi
     = m_P \sqrt{\frac{w}{2}}.
\end{align}
where $\vev{({\cosh\Delta+\cos\phi})^{-1}}_\phi = 1/\sinh\Delta$.
This is consistent with the result obtained in Ref.\cite{Co:2021lkc}
derived assuming PQ conservation.

Altogether, when the PQ field is away from the minimum, i.e. $z\gtrsim 3$,
\begin{align}
S^2 \propto e^{w/2} \propto a^{-3}, \quad
\rho_{{P}} \propto w e^{w/2} \propto a^{-3}, \quad
\dot{\theta} \propto \sqrt{w} \propto a^{0},
\end{align}
while, when the PQ field reaches its minimum value, i.e. $z \sim 0$,
\begin{align}
S^2 \propto e^{w/2} \propto a^0, \quad
\rho_{{P}} \propto w e^{w/2} \propto a^{-6}, \quad
\dot{\theta} \propto \sqrt{w} \propto a^{-3}.
\end{align}
Thus our solution describes both matter oscillation and kination.
The PQ field energy becomes kination-like around
\begin{align}
 u_K := u_1 + \frac{1}{3}\log C_w,
\end{align}
where $z=1$.
Here, $\Delta \to \infty$ at this time is assumed
so that the thermalization completes successfully.
These averaged values are used in the calculations of various observables next section.

\subsection{Energy evolution and thermalization}
\label{sec-thermalization}

The rest of the work needed to numerically calculate the PQ field dynamics
is to determine $\Delta$.
In the minimal KSVZ model, the saxion is dominantly thermalized
by gluon scattering~\cite{Bodeker:2006ij,Laine:2010cq,Mukaida:2012qn}
and the decay to axions~\cite{Co:2016xti}
whose rates are respectively given by
\begin{align}
\Gamma := \Gamma_g + \Gamma_a, \quad
\Gamma_g = b_g \frac{T^3}{S^2},
\quad
\Gamma_a = \frac{m_P^3}{32\pi S^2},
\end{align}
where $b_g = 1.0\times 10^{-5}$ in our numerical analysis.
The evolution equations for the energy densities are given by~\footnote{
In our analysis, the dissipation effect of the axion~\cite{Moroi:2014mqa},
whose rate is given by $\sim b_g m_P^2 T/S^2$, is omitted,
since it might be negligible~\cite{Harigaya:2019qnl}.
}
\begin{align}
\label{eq-deqrhoP}
\dot{\rho}_P + 6 H \abs{\dot{P}}^2
=&\ - \left(\Gamma_g + \Gamma_a \right) \dot{S}^2, \\
\label{eq-deqrhoR}
\dot{\rho}_R + 4 H \rho_R =&\ + \Gamma_g  \dot{S}^2, \\
\label{eq-deqrhoA}
\dot{\rho}_a + 4 H \rho_a =&\ + \Gamma_a  \dot{S}^2,
\end{align}
with
\begin{align}
3 M_p^2 H^2 = \rho_P + \rho_R + \rho_a.
\end{align}
Here, $\rho_a$ is the radiation energy density of the axion produced from the decay.
We used Eqs.~\eqref{eq-deqS} and~\eqref{eq-deqq} to derive Eq.~\eqref{eq-deqrhoP}.
As discussed before, only the kinetic energy of the radial direction
is converted to radiation energy.
The PQ field energy $\rho_P$ is given by Eq.~\eqref{eq-rhoP}.

We numerically solve the evolution equations of the energy densities
together with that of $\Delta$ given by Eq.~\eqref{eq-deqDelta}.
Defining $X_R$ and $X_a$ as
\begin{align}
 \rho_R = \rho_R^1 X_R e^{-4(u-u_1)},
\quad
 \rho_a = \rho_R^1 X_a e^{-4(u-u_1)},
\end{align}
with $\rho^1_R := \pi^2g_* T_1^4/30$.
Averaging the right-hand sides over the rotation,
the equations for $\Delta$, $X_R$ and $X_a$ are given by
\begin{align}
\label{eq-deaDp}
 \Delta^\prime =&\ \frac{e^{-w/2}}{4v_P^2 H}
      \left( b_g X_R^{3/4} T_1^3 e^{-3(u-u_1)} + \frac{m_P^3}{32\pi}\right), \\
\label{eq-deaRp}
 X_R^\prime =&\ \frac{b_g m_P^2 T_1^3}{2\rho_R^1 H} X_R^{3/4} e^{u-u_1}
                \frac{w e^{-\Delta}}{\sinh\Delta},  \\
 X_a^\prime =&\ \frac{m_P^5}{64\pi \rho_R^1 H} e^{4(u-u_1)}
                \frac{w e^{-\Delta}}{\sinh\Delta},
\label{eq-deaAp}
\end{align}
and the initial conditions are
\begin{align}
 \Delta(u_1) = C_\Delta, \quad X_R(u_1) = 1, \quad X_a(u_1) = 0.
\end{align}
Here we use
\begin{align}
 \vev{\frac{\sin^2\phi}{(\cosh\Delta + \cos\phi)^2}}_\phi = \coth\Delta- 1.
\end{align}
We numerically solve Eqs.~\eqref{eq-deaDp}, \eqref{eq-deaRp} and \eqref{eq-deaAp},
so that the evolution of $\Delta$ and the radiation energies are determined.
For the numerical evaluation,
we require that $\Delta(u_K) > 10$ for the completion of the thermalization.

\section{Cosmology}
\label{sec-Cosmo}

\subsection{Thermal history}
\label{sec-therhist}

In this section, we discuss the thermal history in the scenario
based on the analytical solutions.
The numerical results obtained by solving the equations derived in the previous sections
will be shown later.
Since the initial amplitude is large,
the PQ field may dominate the universe at some time.
When the MD starts, $\rho_P = \rho_R$,
the temperature is given by\footnote{We used
\begin{align}
\rho_P(u_M) \sim m_P^2 v_P^2 w_M  e^{w_M/2} (1+e^{-2\Delta})
       = \sqrt{\frac{w_M}{2}} m_P n_\PQ \coth\Delta \notag
\end{align}
with assuming $w_M := w(u_M) \gg 1$.
}
\begin{align}
 T_M \sim&\ \frac{4}{3} \sqrt{\frac{w_M}{2}} \coth C_\Delta \cdot m_P Y_\osc  \\ \notag
    \sim&\ 9\times 10^{7}~\GeV \times \left(\frac{\sqrt{w_M} \coth C_\Delta}{10}\right)
                                   \left(\frac{m_P}{10^6~\GeV}\right)
                                   \left(\frac{Y_\osc}{10}\right),
\end{align}
There is no MD era
if this temperature is lower than the temperature when
the PQ field energy becomes kination-like, i.e.
\begin{align}
\label{eq-TMTK}
 T_M < T_i e^{-u_K}
     =&\ \left(\frac{45v_P^2m_P}{\sqrt{2}\pi^2 g_* Y_\osc}\right)^{\frac{1}{3}}  \\ \notag
       \sim&\ 2\times 10^6~\GeV \times \left(\frac{v_P}{10^8~\GeV}\right)^{\frac{2}{3}}
                                       \left(\frac{m_P}{1~\PeV}\right)^{\frac{1}{3}}
                                       \left(\frac{10}{Y_\osc}\right)^{\frac{1}{3}}.
\end{align}
The MD starts at some time if this condition is not satisfied.
Once the PQ field dominates the energy density of the universe,
it should be thermalized predominantly by gluon scattering.
If, instead, the PQ field is thermalized by the decay to axions,
it contributes to the effective number of neutrinos, $\Delta N_\eff$,
which is severely constrained by current observation.
Thus we assume that the thermalization is dominated by gluon scattering
in the following analytical estimation.

We can understand the thermalization process by solving Eqs.~\eqref{eq-deaDp}
and~\eqref{eq-deaRp} with neglecting the decay to axions.
Assuming the thermalization rate is negligible at $u=u_M$,
$\Delta^\prime \ll 1$, so $\Delta \simeq C_\Delta$ is the constant at this time.
The solution for the $X_R$ when $\Delta^\prime \ll 1$ is given by
\begin{align}
 X_R^{1/4} = 1 + \beta \left(e^{\frac{5}{2}(u-u_M)} - 1\right),
\quad
\beta :=
    \frac{9\sqrt{10}b_g w_M^\frac{1}{4}w_N^{\frac{3}{4}} e^{-C_\Delta}}
         {2\pi^3 g_*^{3/2} \sinh C_\Delta} \frac{m_P^2 M_p}{T_M^3}.
\end{align}
In the analytical analysis, we assume that the thermalization is
ineffective when the MD starts, and thus $\beta \ll 1$.
If this is not true, the discussion in this section is not applicable,
and the numerical evaluation is necessary.
Hence, the thermalization becomes effective
and the non-adiabatic (NA) era starts at $u=u_N$ with
\begin{align}
 u_N = u_M - \frac{2}{5} \log \beta, \label{eq:uNuM}
\end{align}
where $\beta e^{5(u_N-u_M)/2} = 1$.
This is about a time when the NA era starts,
because the evolution is adiabatic while $X_R = 1$
and $X_R^{1/4} \sim \beta e^{5/2(u-u_M)} >1$ at $u > u_N$.
$u_N < u_K$ is necessary for successful thermalization,
since the thermalization rate drops faster than the Hubble constant at $u > u_K$.
At $u\gtrsim u_N$, the scaling of the temperature
becomes $T\propto X_R^{1/4} e^{-u} \propto e^{3/2u}$.

We can estimate the scale when the thermalization completes, $u_\ther$,
by equating $\rho_{\dot{S}}$ without the thermalization effect
and the radiation energy in the NA era,
\begin{align}
\left. \rho_{\dot{S}} (u_\ther) \right|_{\Delta = C_\Delta}
 = \rho_R(u_\ther) \sim \rho_R(u_M) e^{-4(u_N-u_M)} e^{6(u_\ther - u_N)}. \label{eq:Tth}
\end{align}
This equation means that the kinetic energy stored in the radial direction
converts to radiation energy.
In the second equality,
we assume the scaling of the temperature $T \propto e^{-u}$ ($e^{3u/2}$)
before (after) $u = u_N$.
From this relation, we obtain the thermalization temperature as follows.
The left-hand side of Eq.~\eqref{eq:Tth} is given by
\begin{align}
\label{eq-eqth}
\left. \rho_{\dot{S}}(u_\ther) \right|_{\Delta = C_\Delta}
= \sqrt{\frac{w_\ther}{2}} m_P n_\PQ(u_M) e^{-3(u_\ther-u_N)} e^{-3(u_N-u_M)}
                  \frac{e^{-C_\Delta}}{2\sinh C_\Delta},
\end{align}
with $w_\ther = w(u_\ther)$.
While the right-hand side of Eq.~\eqref{eq:Tth} is
\begin{align}
 \rho_R(u_M) e^{-4(u_N-u_M)}e^{6(u_\ther-u_N)}
 =&\ \rho_P(u_M) e^{-4(u_N-u_M)}e^{6(u_\ther-u_N)}  \\ \notag
 \sim&\ \sqrt{\frac{w_M}{2}} m_P n_\PQ(u_M)
        \coth C_\Delta e^{-4(u_N-u_M)}e^{6(u_\ther-u_N)}.
\end{align}
Equating them, we get
\begin{align}
e^{9(u_\ther-u_N)} = \sqrt{\frac{w_\ther}{w_M}}
                    \frac{e^{-C_\Delta}}{2\cosh C_\Delta} e^{u_N-u_M}.
\end{align}
Then given the scaling laws of the temperature assumed in Eq.~\eqref{eq-eqth},
this becomes
\begin{align}
\label{eq:scalings}
 \left(\frac{T_\ther}{T_N}\right)^6 =
 \sqrt{\frac{w_\ther}{w_M}}
                    \frac{e^{-C_\Delta}}{2\cosh C_\Delta} \frac{T_M}{T_N}.
\end{align}
Thus,
we arrive at
\begin{align}
 T_\ther^6 = \sqrt{\frac{w_\ther}{w_M}} \frac{e^{-C_\Delta}}{2\cosh C_\Delta} T_M T_N^5.
\end{align}
Finally, using Eq.~\eqref{eq:uNuM}, we obtain
\begin{align}
\label{eq-TthApp}
 T_\ther =&\ \left(\sqrt{\frac{w_\ther}{w_M}} \frac{e^{-C_\Delta}}{2\cosh C_\Delta}
                  \beta^2 \right)^{\frac{1}{6}} T_M  \\ \notag
         \sim&\ 4\times 10^7~\GeV\times
              \left( \frac{\sqrt{w_\ther w_N^3}}{700}
                        \frac{ e^{-3C_\Delta} \tanh C_\Delta}
                             {\sinh^2 C_\Delta} \right)^{\frac{1}{6}}
             \left(\frac{b_g}{10^{-5}}\right)^{\frac{1}{3}}
             \left(\frac{228.75}{g_*}\right)^{\frac{1}{2}}
             \left(\frac{m_P}{10^6~\GeV}\right)^{\frac{2}{3}}.
\end{align}
The dilution factor is given by
\begin{align}
 D:= \frac{s_{\mathrm{aft}}}{s_{\mathrm{bef}}}
= \left.
\left( \frac{\rho_R(u)}{\rho_R^1 e^{-4(u-u_1)}}\right)^{\frac{3}{4}}\right|_{u\gg u_\ther}
   = X_R(u\gg u_\ther)^{\frac{3}{4}},
\end{align}
with $s_\mathrm{aft}$ ($s_\mathrm{bef}$) the entropy after (before) dilution
by the thermalization.
In the numerical evaluation, we evaluate this directly at $u=u_K+2$.
This is estimated as
\begin{align}
 D\sim&\ \frac{T_\ther^3}{T_M^3 e^{-3(u_\ther-u_M)}}
 = \frac{T_\ther^5}{T_N^5}   =
           \sqrt{\frac{w_\ther}{w_M}}\frac{e^{-C_\Delta}}{{2\cosh C_\Delta}}
                \frac{T_M}{T_\ther} \\ \notag
 \sim&\ 10\times \left(\frac{w_\ther^{5/2} e^{-3C_\Delta}\cosh C_\Delta }
                           {10^5 \times w_N^{3/2} \sinh^4 C_\Delta }\right)^{\frac{1}{6}}
                   \left(\frac{g_*}{228.75}\right)^{\frac{1}{2}}
                   \left(\frac{10^{-5}}{b_g}\right)^{\frac{1}{3}}
                   \left(\frac{Y_\osc}{50}\right)
                   \left(\frac{m_P}{10^6~\GeV}\right)^{\frac{1}{3}},
\end{align}
where the scalings of $T$ used to derive Eq.~\eqref{eq-eqth} are assumed again.
Thus the dilution factor is about $\order{10}$ for $Y_\osc \sim \order{10}$.
After the successful thermalization,
the PQ field reaches its minimum value when the temperature is
\begin{align}
 T_K := T(u_K) \sim
  \left(\frac{45v_P^2 m_P}{\sqrt{2}\pi^2 g_* Y_\PQ}\right)^{\frac{1}{3}},
\end{align}
where $Y_\PQ$ is the PQ number yield after the dilution.
The value of $T_K$ is given by in the right-hand side of Eq.~\eqref{eq-TMTK}
by replacing $Y_\osc \to Y_\PQ$.
The kination domination (KD) era exists if $\rho_P(u_K) > \rho_R(u_K)$.
This condition is approximately given by
\begin{align}
1 >  \frac{1215 v_P^2}{32\pi^2 g_* \Omega^3 Y_\PQ^4 m_P^2}
 \sim 3 \times   \left(\frac{5}{Y_\PQ}\right)^4
                 \left(\frac{228.75}{g_*}\right)
                 \left(\frac{v_P}{10^8~\GeV}\right)^2
                 \left(\frac{10^6~\GeV}{m_P}\right)^2,
\end{align}
where $\Omega := e^{\Wcal(1)/2}(\Wcal(1)-1)+1 \sim 0.43$.
The KD era is absent for typical values of the parameters,
although it could happen for a certain parameter set.
No KD means that radiation becomes the dominant form of energy
after thermalization.

The axion contribution to the effective number of neutrinos is given by~\cite{Marsh:2015xka}
\begin{align}
 \Delta N_\eff =
 \frac{43}{7}\left(\frac{10.75}{g_*(T_{\mathrm{th}})}\right)^{\frac{1}{3}}
              \left.\frac{\rho_a}{\rho_R}\right|_{u\gg u_\ther}.
\end{align}
This is estimated as
\begin{align}
 \Delta N_\eff \simeq&\
   \frac{43}{7}\left(\frac{10.75}{g_*(T_\ther)}\right)^{\frac{1}{3}}
  \left. \frac{\Gamma_{a}}{\Gamma_{g }} \right|_{T=T_\ther}
=     \frac{43}{7} \left(\frac{10.75}{g_*(T_{\ther})}\right)^{\frac{1}{3}}
        \frac{m_P^3}{32\pi b_g T_\ther^3}  \\ \notag
 =&\ 0.28 \times \left(\frac{228.75}{g_*(T_\ther)}\right)^{\frac{1}{3}}
     \left(\frac{10^{-5}}{b_g}\right)
     \left(\frac{m_P/T_\ther}{0.05}\right)^3.
\end{align}
Thus, $\Delta N_\eff$ will be suppressed
if $T_\ther$ is an order of magnitude larger than $m_P$.
The current limit is $\Delta N_\eff < 0.3$ ~\cite{Aghanim:2018eyx}.

\begin{figure}[t]
\centering
\begin{minipage}[t]{0.49\textwidth}
\centering
\includegraphics[height=90mm] {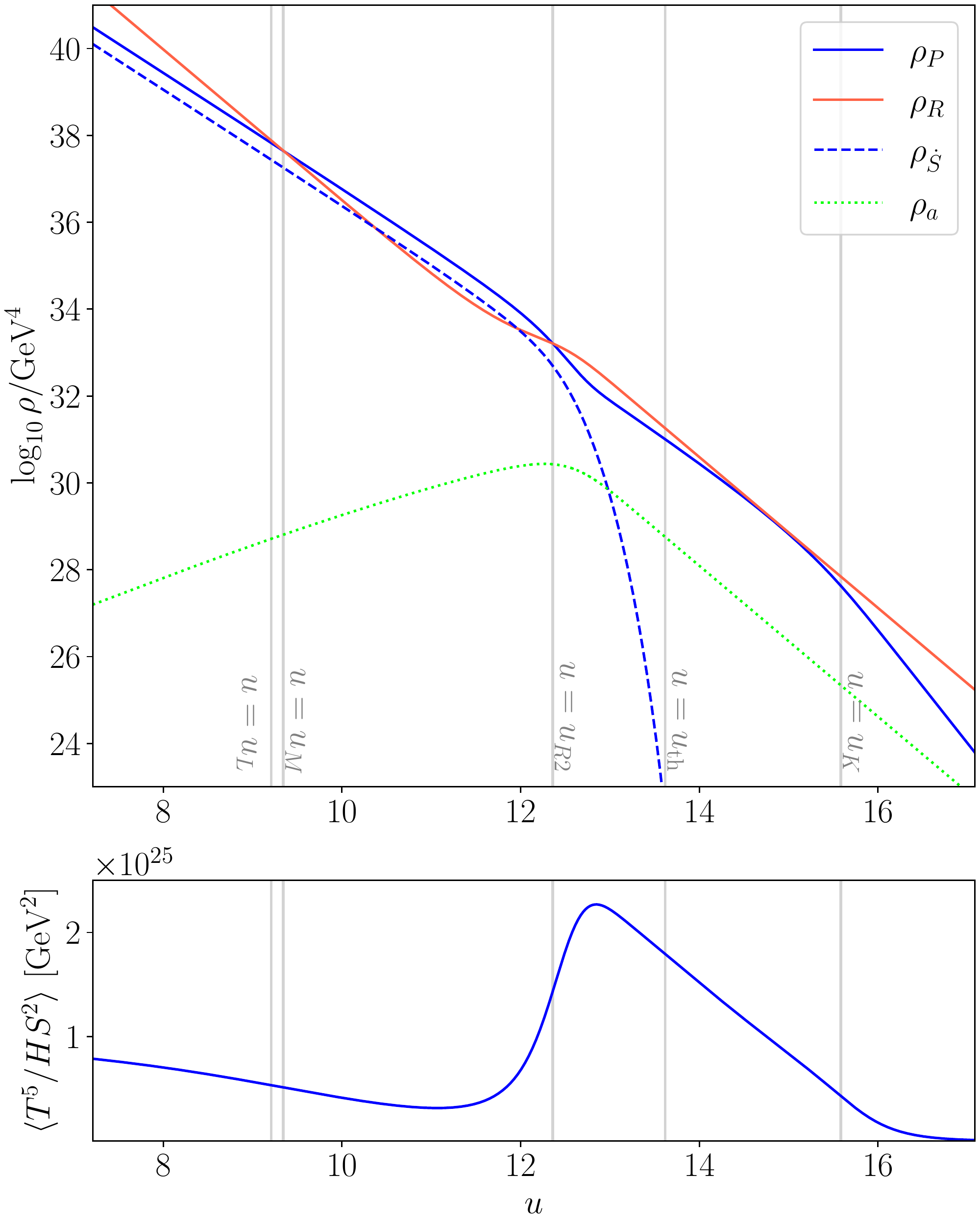}
\end{minipage}
\begin{minipage}[t]{0.49\textwidth}
\centering
\includegraphics[height=90mm] {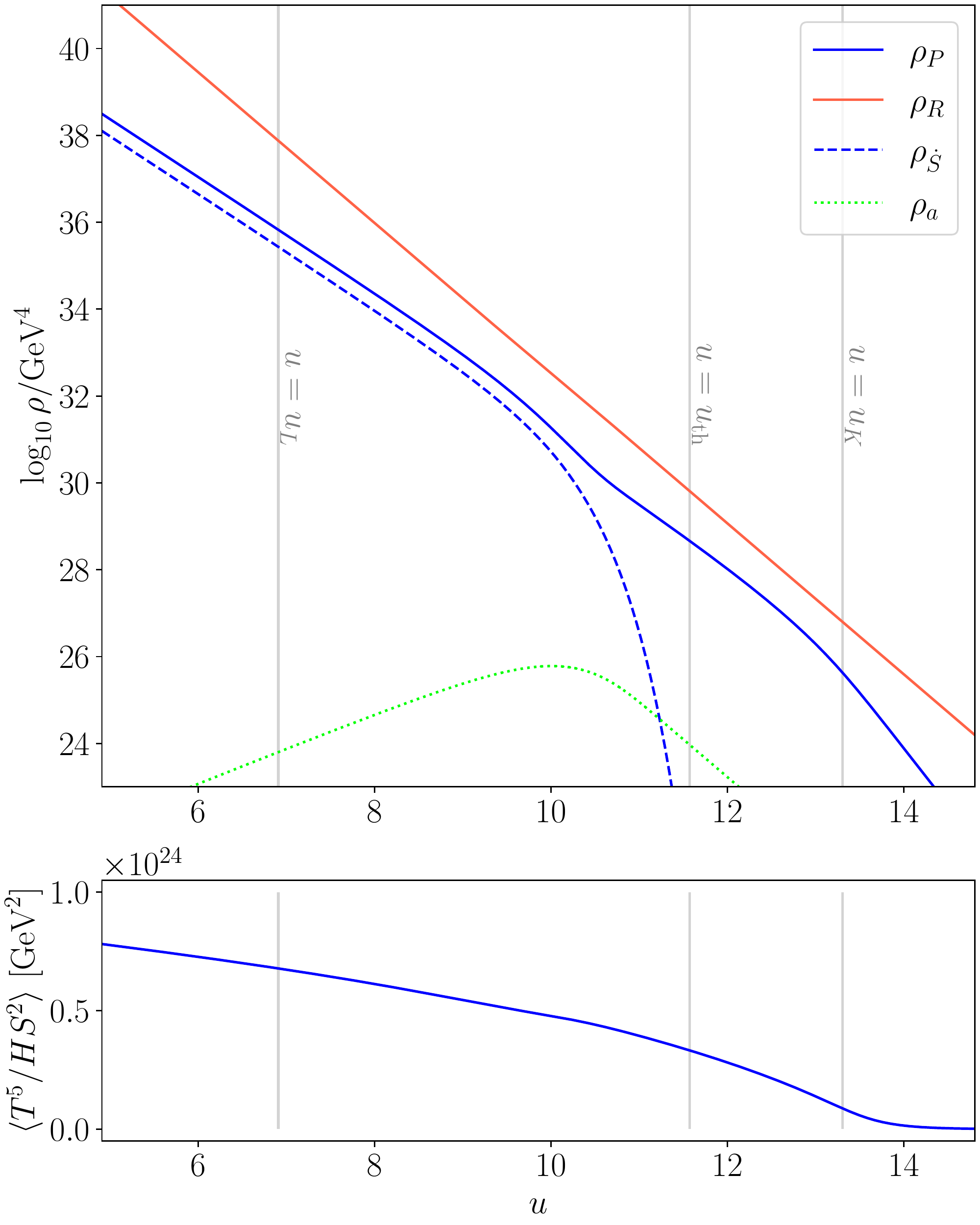}
\end{minipage}
\caption{\label{fig-evolrho}
The evolution of the energy densities of the PQ field (blue solid), radiation (red),
radial direction (blue dashed) and axion (green dotted).
$n=10$, $\theta_i=\pi/20$, $T_i = 10^{13}$ GeV, $v_P = 10^8$ GeV and $m_P = 10^6$ GeV
in the left panel,
and
$n=8$, $\theta_i=\pi/16$, $T_i = 10^{12}$ GeV, $v_P = 10^{8}$ GeV and $m_P = 10^5$ GeV
in the right  panel.
The lower panels show $\vev{T^5/HS^2}$ which is integrated to calculate
the baryon asymmetry.
}
\end{figure}
Figure~\ref{fig-evolrho} shows the evolution of energy densities
at the benchmark points~\footnote{
Note, $u_N$
is not shown in the figure since it is not calculated in the numerical analysis.
$u_N$ is introduced for the analytical estimation.
}.
The evolution equation of $\Delta$ and the energy densities are given
by Eqs.~\eqref{eq-deqrhoP},~\eqref{eq-deqrhoR} and~\eqref{eq-deqrhoA}.
In the figure,
$u_L$ is the scale when the lepton number violating interaction
becomes out of equilibrium as discussed in the next section
and $u_{R2}$ is the scale when the second RD starts after the MD starts.
On the left panel,
$n=10$, $\theta_i=\pi/20$, $T_i = 10^{13}$ GeV, $v_P = 10^8$ GeV and $m_P = 10^6$ GeV.
The MD starts at $u\sim 9.3$, then the thermalization starts at $u\sim 11$
and it ends at $u \sim 13.5$.
The dilution factor is $6.0$ at this point.
We see that the kinetic energy of the radial direction, $\rho_{\dot{S}}$,
is converted to the radiation energy through the thermalization process,
so only a fraction of the PQ field energy is lost.
The PQ field energy becomes kination-like at $u\sim 15.8$,
but the dominant energy at this time is the radiation energy,
since the radiation energy starts to dominate during the thermalization process.
On the right panel,
$n=8$, $\theta_i=\pi/16$, $T_i = 10^{12}$ GeV, $v_P = 10^8$ GeV and $m_P = 10^5$ GeV.
In this case, the radiation energy dominates the universe throughout the history
because of the smaller amplitude of the PQ field due to smaller $n$.

\begin{figure}[t]
\centering
\includegraphics[height=0.86\textwidth] {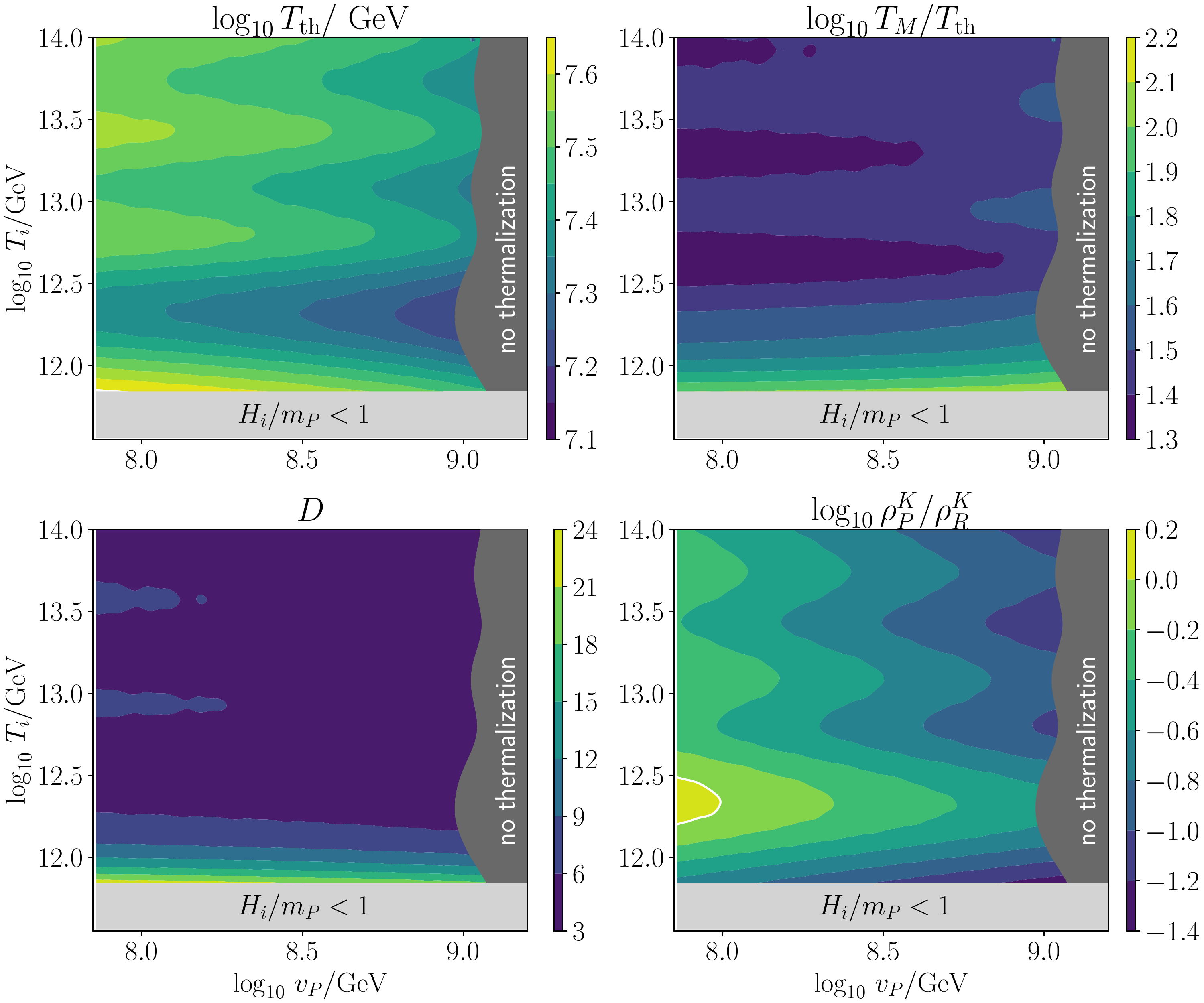}
\caption{\label{fig-Tratios}
The values of $T_\ther$ (top-left), $T_M/T_\ther$ (top-right),
$D$ (bottom-left) and $\rho_P^K/\rho_R^K$ (bottom-right)
in the case of $n=10$, $\theta_i=\pi/20$, and $m_P = 10^6~\GeV$.
}
\end{figure}
Figure~\ref{fig-Tratios}
shows the values of $T_\ther$ (top-left), $T_M/T_\ther$ (top-right),
$D$ (bottom-left) and $\rho_P^K/\rho_R^K$ (bottom-right)
calculated by the numerical calculation on the $(v_P, T_i)$ plane
when $n=10$, $\theta_i=\pi/20$, and $m_P = 10^6~\GeV$.
From the top-left panel,
the thermalization temperature, where $\Delta = 10$,
is $\order{10^7~\GeV}$ as expected from our estimation Eq.~\eqref{eq-TthApp}.
It should be noted that this is the temperature when the thermalization completes,
and the thermalization itself occurs slightly before this time.
We can see this from Fig.~\ref{fig-evolrho},
the radiation energy is non-adiabatic at $11.5 \lesssim u \lesssim 13$
and is before $u = u_\ther \sim 13.5$.
Hence, e.g. $\Delta N_\eff$ is dominantly produced around $u \lesssim 13$
and the typical temperature is effectively higher than $T_\ther$.
The thermalization does not complete
in the dark gray region at $v_P \gtrsim 10^{9}~\GeV$,
i.e. $\Delta(u_K) < 10$ in this region.
The PQ field reaches its minimum value too early, before thermalization completes,
if $v_P$ is too large.
In the top-right panel, the ratio of $T_M$ to $T_\ther$ is shown.
The ratio is at most $\order{100}$ for $T_i \gtrsim 10^{12}~\GeV$,
and, in this case, the MD era does not last such a long time.
The dilution factor $D$ is shown in the bottom-left panel.
$D \sim \order{10}$ reflecting the short MD era at $T \gtrsim 10^{12}~\GeV$,
while it can be larger for lower $T_i$.
The ratio of the PQ field to radiation energy at $u=u_K$ is shown
in the bottom-right panel.
The PQ field energy is smaller than the radiation energy throughout the figure,
except for the small region where $v_P \lesssim 10^8~\GeV$ and $T_i \sim 10^{12.3}~\GeV$.
Although it is not the leading energy,
the kination energy and the radiation energy are comparable
at $T_i \gtrsim 10^{12}~\GeV$.
This is because the ratio $T_M/T_\ther$ is small and the PQ field and radiation
energies are still comparable when thermalization occurs.

In our numerical analysis, we focus on the two parameter sets,
\begin{itemize}
 \item {$n=10$, $\theta_i = \pi/20$, $m_P=10^6~\GeV$ }
 \item {$n=8$, $\theta_i = \pi/16$, $m_P=10^5~\GeV$ }
\end{itemize}
with varying $v_P$ and $T_i$.
We name the former (latter) scenario as $n=10$ ($n=8$) scenario.
In both cases, we take $c_H = \la = 1$ and $A_P = m_P$.

\subsection{Baryon asymmetry}
\label{sec-Basym}

The PQ number density produced from the rotational motion
is converted to the baryon asymmetry via sphaleron processes
and MSSM interactions~\cite{Co:2019wyp,Co:2020jtv}.
In the lepto-axiogenesis scenario,
the generated lepton chiral asymmetry and Higgs number asymmetries are
converted into the $\BL$ asymmetry via the Weinberg operator $(LH_u)^2$,
which can be obtained by integrating out
Majorana neutrinos in the type-I see-saw mechanism~\cite{Yanagida:1979as,Gell-Mann:1979vob,Minkowski:1977sc,Mohapatra:1979ia}.
The $\BL$ asymmetry is finally converted to the baryon asymmetry
via the weak sphaleron process.
The Boltzmann equation for the $\BL$ asymmetry reads
\begin{align}
\label{eq-BoltzBL}
\dot{n}_\BL + 3 H n_\BL = c_\BL \dot{\theta} T^2 \Gamma_L.
\end{align}
Here, $\Gamma_L$ is a rate of conversion from the Higgs asymmetry and/or lepton asymmetry
to $\BL$ asymmetry via the Weinberg operator and is given by
\begin{align}
\label{eq-GamL}
 \Gamma_L \simeq \frac{1}{4\pi^3} \frac{m_\nu^2}{v_H^4} T^3.
\end{align}
$m_\nu^2 := \sum_{i=1,2,3} m_{\nu_i}^2$
is the sum of neutrino masses squared.
This should be in the range $0.0025~\eV^2 \lesssim m_\nu^2 \lesssim 0.03~\eV^2$
to be consistent
with the neutrino oscillation data~\cite{Esteban:2020cvm,Capozzi:2018ubv,deSalas:2017kay}
and
the cosmological bound $\sum m_{\nu_i} < 0.26~\eV$~\cite{Aghanim:2018eyx}.
The constant $c_\BL$ depends on the scattering rates of MSSM particles.
The baryon asymmetry per entropy density, $Y_B := n_B/s$, is given by
\begin{align}
\label{eq-IntnB}
Y_B  = Y_\PQ  \times \frac{c_Bm_\nu^2}{4\pi^3v_H^4}
                 \int^{u}_{u_L} du^\prime \frac{T^5}{H S^2}
\end{align}
where $c_B := 10/31 \times c_\BL$ in the MSSM~\cite{Harvey:1990qw}.
Here, $u_L$ is when the lepton number violating process goes out of equilibrium.
The value of $c_B$ is $\order{0.01-0.1}$ depending on the details
of the couplings~\cite{Co:2020jtv}.
Hence, we consider $3\times 10^{-5}~\eV^2 < c_B m_\nu^2 <  3\times 10^{-3}~\eV^2$
as a suitable range.
In our analysis, we assume that the Yukawa interactions,
$y_\nu N_1 LH_u$ , are in thermal equilibrium when $T > M_{N_1}$,
with $M_{N_1}$ the mass of the lightest Majorana neutrino,
as in the strong washout regime
of thermal leptogenesis~\cite{Fukugita:1986hr,Davidson:2008bu}.
This is particularly motivated by the Yukawa texture,
$Y_\nu \sim Y_u$,
in Grand Unified Theories~\cite{Poh:2015wta,Poh:2017xvg,Raby:2017ucc},
where $Y_u$ is the Yukawa coupling matrix for up quarks.
We shall take $M_{N_1} = 10^9~\GeV$ for concreteness,
so $u_L = \log (T_i/10^9~\GeV)$~\footnote{With this assumption,
the strong sphaleron is always in equilibrium at $u > u_L$.
}.
The baryon asymmetry is evaluated
by numerically integrating the averaged function, $\vev{T^5/HS^2}$,
from $u_L$ to $u_K + 2$.
The evolution of $\vev{T^5/HS^2}$ is shown in the lower panels of Fig.~\ref{fig-evolrho}.

Let us estimate the baryon asymmetry analytically.
If there is no MD,
the integration can be done with neglecting the variation
of $\Delta$ in $w$~\footnote{
We use the integration formula
\begin{align}
\notag
 \int dx \sqrt{\Wcal\left(a e^{-bx}\right)} = -\frac{2\sqrt{w}}{3b} \left(3+w\right),
\quad
w = \Wcal\left(a e^{-bx}\right).
\end{align}
with $a,b$ constants.
}
\begin{align}
\label{eq-IntRD}
 \int^u_{u_L} du^\prime \vev{\frac{T^5}{H S^2}}
 \sim \left(\frac{90}{\pi^2g_*}\right)^\frac{3}{2} \frac{m_P M_p}{36 Y_\PQ}
   \left( \sqrt{\frac{w_L}{2}} (3+w_L) - \sqrt{\frac{w(u)}{2}} (3+w(u)) \right),
\end{align}
where $w_L := w(u_L)$.
Since $w(u) \ll w_L$, at sufficiently late times,
the baryon asymmetry is given by
\begin{align}
\label{eq-YBRD}
 Y_B^\RD \sim&\ \frac{c_B m_\nu^2 m_P M_p}{144\pi^3v_H^4}
            \left(\frac{90}{\pi^2g_*}\right)^\frac{3}{2}\sqrt{\frac{w_L}{2}} (3+w_L)
\\ \notag
\sim&\
 1\times 10^{-10} \times \left(\frac{c_B m_\nu^2}{0.003~\eV^2}\right)
                               \left(\dfrac{m_P}{10^5~\GeV}\right),
\end{align}
where $w_L = 20$ is assumed in the second equality.
The observed value, $Y_B^{\mathrm{obs}} \sim 8.7\times 10^{-11}$~\cite{Aghanim:2018eyx},
can be explained for $m_P \gtrsim \order{10^5~\GeV}$.

If the MD exists,
the integration during the period of RD is given by Eq.~\eqref{eq-IntRD}
with formally replacing $u \to u_M$.  The second term may not be negligible in this case.
This contribution is diluted by the entropy production from the thermalization.
Neglecting $\order{1}$ variations of $w$ and $\tanh\Delta$,
$\vev{T^5/HS^2}$ scales as $e^{12u}$ and $e^{-u/2}$
during the adiabatic and the NA era, respectively~\footnote{
The scaling-low becomes $e^{15/2u}$ if the RD starts during the thermalization.
The asymmetry is produced at the end of thermalization also in this case,
and hence the estimation here is not changed.
}.
Hence, the baryon asymmetry is predominantly produced when the thermalization ends.
We can see this from the lower-left panel of Fig.~\ref{fig-evolrho}.
Thus the baryon asymmetry produced after an epoch of MD is estimated as
\begin{align}
\label{eq-IntMD}
 Y_B^\MD \sim&\ Y_\PQ \frac{c_B m_\nu^2}{4\pi^3v_H^4} \frac{T_\ther^5}{H_\ther S_\ther^2}
         \sim \frac{c_B m_\nu^2 M_p}{32\pi^3v_H^4}
              \left(\frac{90}{\pi^2g_*}\right)^\frac{3}{2}
              \left(\frac{w_\ther}{2}\right)^{\frac{1}{4}}
              \sqrt{\frac{3m_P T_\ther}{Y_\PQ}} \\ \notag
   \sim&\ 1\times 10^{-10} \times
           \left(\frac{c_B m_\nu^2}{0.003~\eV^2}\right)
           \left(\frac{w_\ther}{5}\right)^{\frac{1}{4}}
           \left(\frac{m_P}{10^6~\GeV}\right)^{\frac{1}{2}}
           \left(\frac{T_\ther}{10^7~\GeV}\right)^{\frac{1}{2}}
           \left(\frac{10}{Y_\PQ}\right)^{\frac{1}{2}}.
\end{align}
Thus, in this case, $m_P \gtrsim \order{10^6~\GeV}$ may be necessary to explain the baryon asymmetry.

\begin{figure}[t]
\centering
\begin{minipage}[t]{0.495\textwidth}
\centering
\includegraphics[height=74mm] {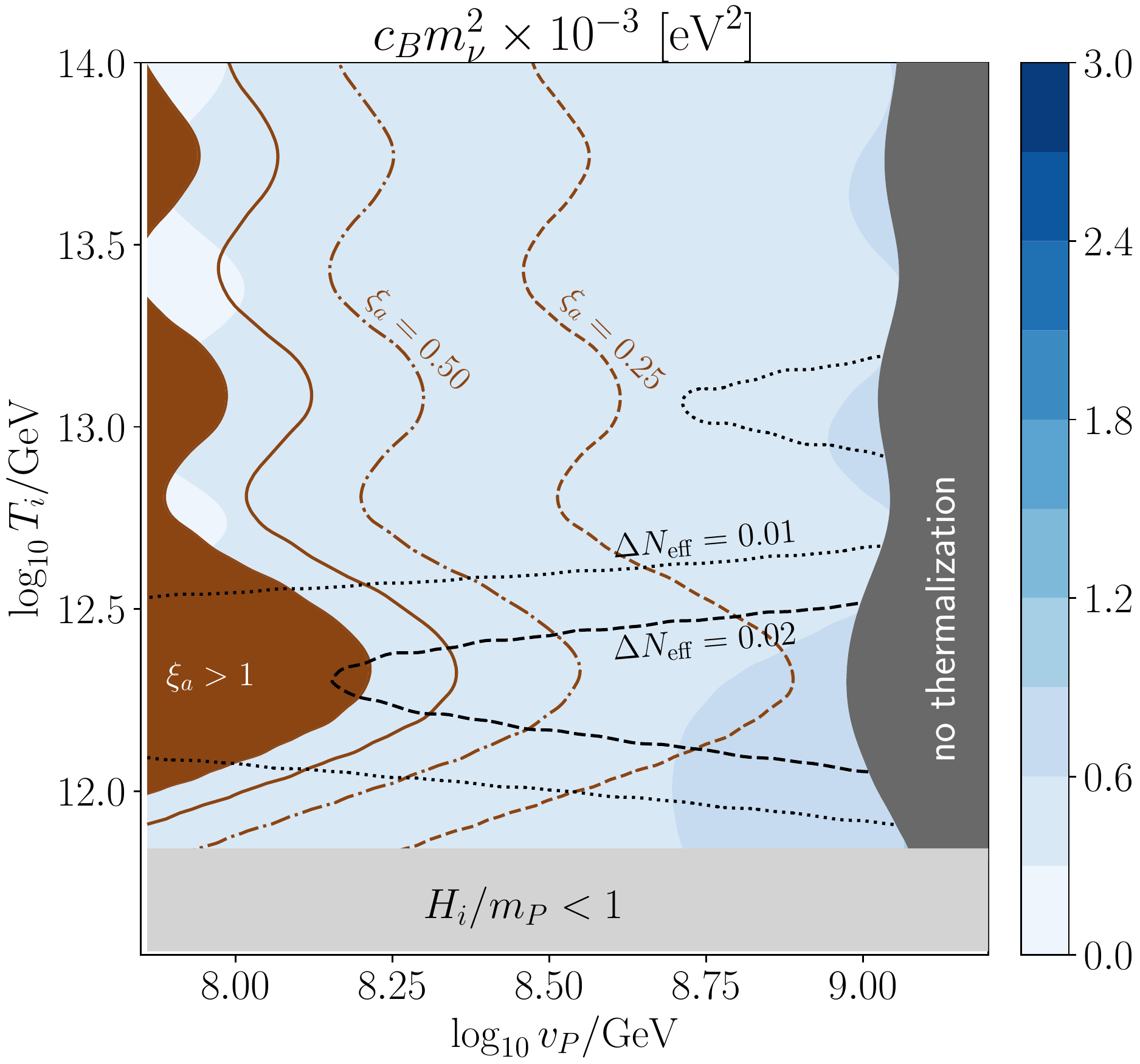}
\end{minipage}
\begin{minipage}[t]{0.495\textwidth}
\centering
\includegraphics[height=74mm] {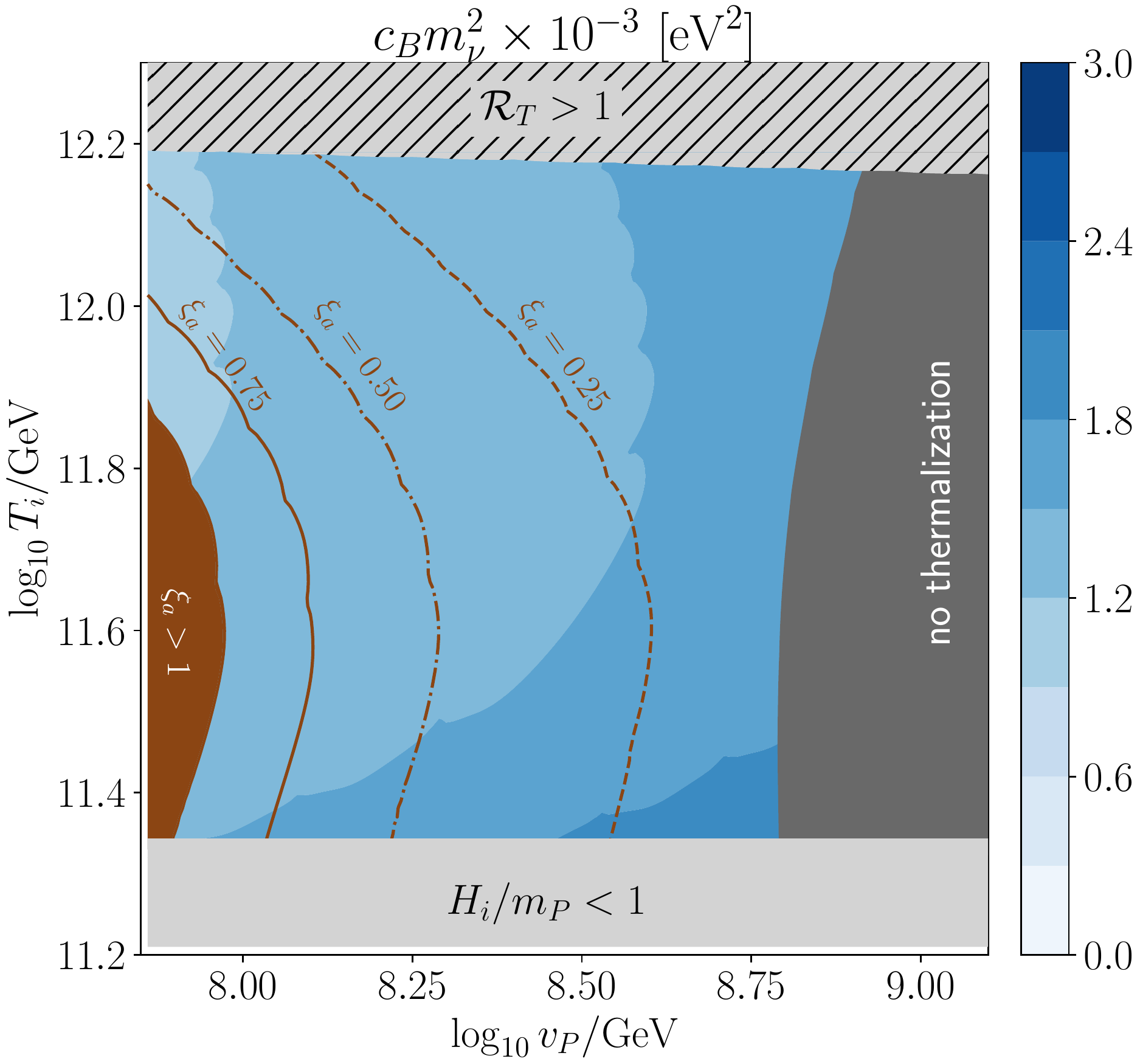}
\end{minipage}
\caption{\label{fig-cmn}
The values of $c_B m_\nu^2$ to explain the baryon asymmetry
in the $n=10$, $m_P=10^{6}~\GeV$ ($n=8$, $m_P=10^{5}~\GeV$) scenario
on the left (right) panel.
The thermalization does not complete in the dark gray region.
$\Delta N_\eff = 0.02$ ($0.01$) on the black dashed (dotted) line.
The fraction of the axion density, $\xi_a := \Omega_a/\Omega_\DM$,
is $0.75, 0.50, 0.25$ on the brown solid, dot-dashed, dashed lines, respectively,
when $N_\DW = 1$. The axion is overproduced in the brown region.
The relative importance of the thermal-log term to the mass term,
$\mathcal{R}_T$ defined in Eq.~\eqref{eq-defRT},
is larger than unity in the gray hatched region.
}
\end{figure}

Figure~\ref{fig-cmn} shows values of $c_B m_\nu^2$
to explain the baryon asymmetry based on the numerical calculation.
The meanings of the regions and lines are explained in the caption.
The result of the $n=10$ ($n=8$) scenario is shown in the left (right) panel.
In the $n=10$ scenario, the MD era exists at some time,
and hence the baryon asymmetry is produced predominantly at the end of thermalization.
The baryon asymmetry is explained at 
$v_P \lesssim 10^{9.2}~\GeV$
with $c_B m_\nu^2 \sim (3\mathrm{-}12)\times 10^{-4}~\eV^2$.
On the black dashed (dotted) line, $\Delta N_\eff = 0.02$ ($0.01$)
and hence inside the dashed line
is accessible in future observations~\cite{CMB-S4:2016ple}.
The brown lines are the fraction of the axion density in the case of $N_\DW = 1$.
The calculation of axion density is shown in the next section.
For $N_\DW > 1$, the density is divided by $N_\DW$.
In the $n=8$ scenario,
there is no MD era anywhere in the parameter space
since the amplitude of the PQ field is smaller due to the smaller value of $n$.
We note that $v_P \gtrsim \order{10^9~\GeV}$
may cause the axion quality problem from Eq.~\eqref{eq-DeltaTheta},
but this region is already incompatible with successful thermalization.
The smaller $m_P = 10^5~\GeV$ is enough to explain the baryon asymmetry
due to the absence of the MD era.
The relative importance of the thermal-log term to the mass term,
$\mathcal{R}_T$ defined in Eq.~\eqref{eq-defRT},
is larger than unity in the gray hatched region.
We need to take the thermal-log effect at $u>u_1$ into account in this region of parameter space
for a reliable calculation.

\subsection{Dark matter density}
\label{sec-DM}

In this model, the axion $\phi$ and the LSP $\chi$ contribute
to the DM density, i.e.
\begin{align}
  \Omega_\DM = \Omega_a + \Omega_\chi =: (\xi_a + \xi_\chi) \Omega_\DM.
\end{align}
We assume that the observed DM density,
$\Omega_\DM^\obs h^2 = 0.12$~\cite{Aghanim:2018eyx},
is explained by these two particles.
Since the reheating temperature $T_i$ is large for $H_i > m_P$, 
the LSP will be predominantly produced by the decay of gravitinos.
We shall discuss the conditions necessary to explain the DM density
and their consequences for DM searches; especially for indirect detection.

\subsubsection{Axion density}

\begin{figure}[t]
\centering
\begin{minipage}[t]{0.48\textwidth}
\centering
\includegraphics[height=65mm] {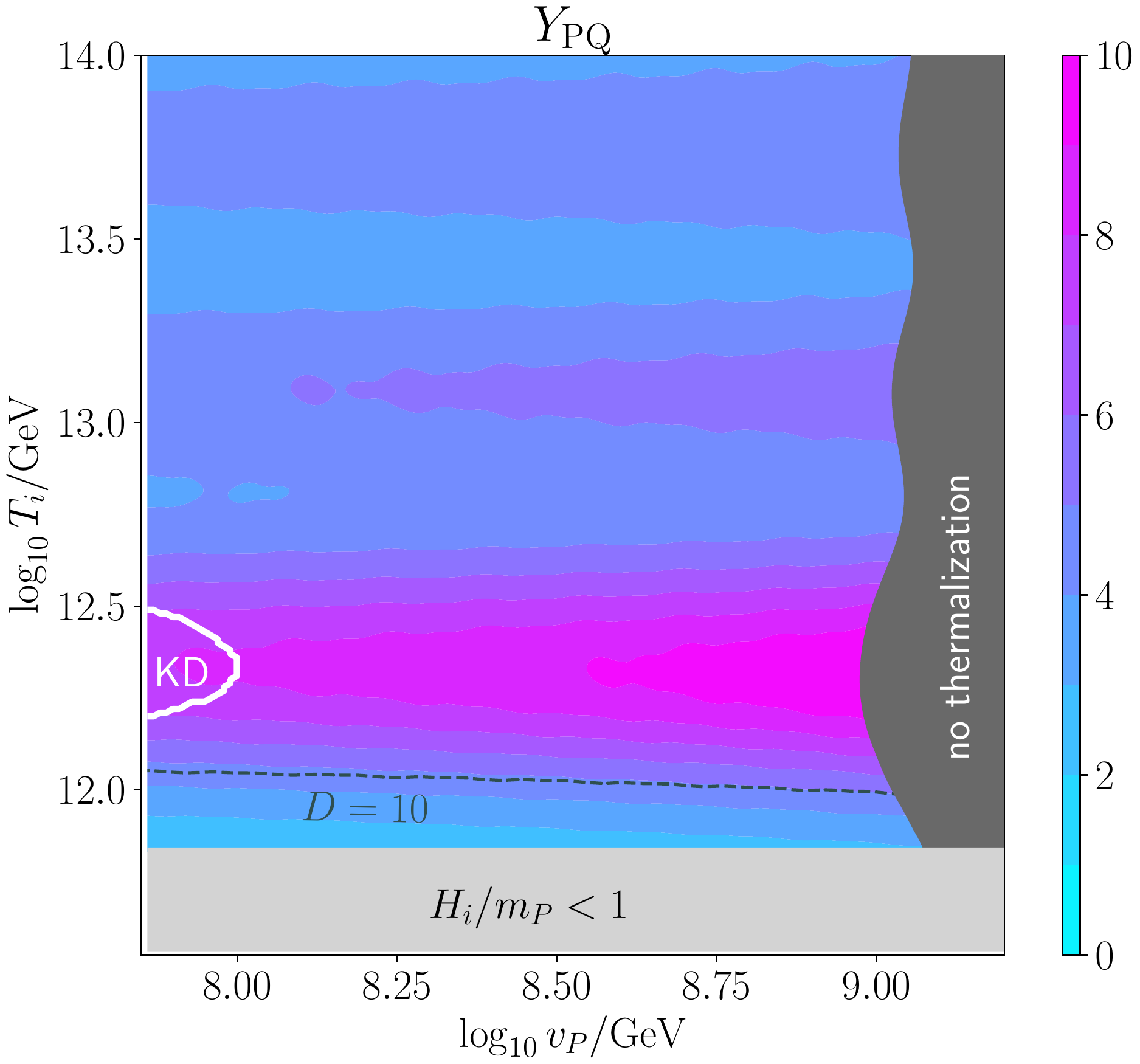}
\end{minipage}
\begin{minipage}[t]{0.48\textwidth}
\centering
\includegraphics[height=65mm] {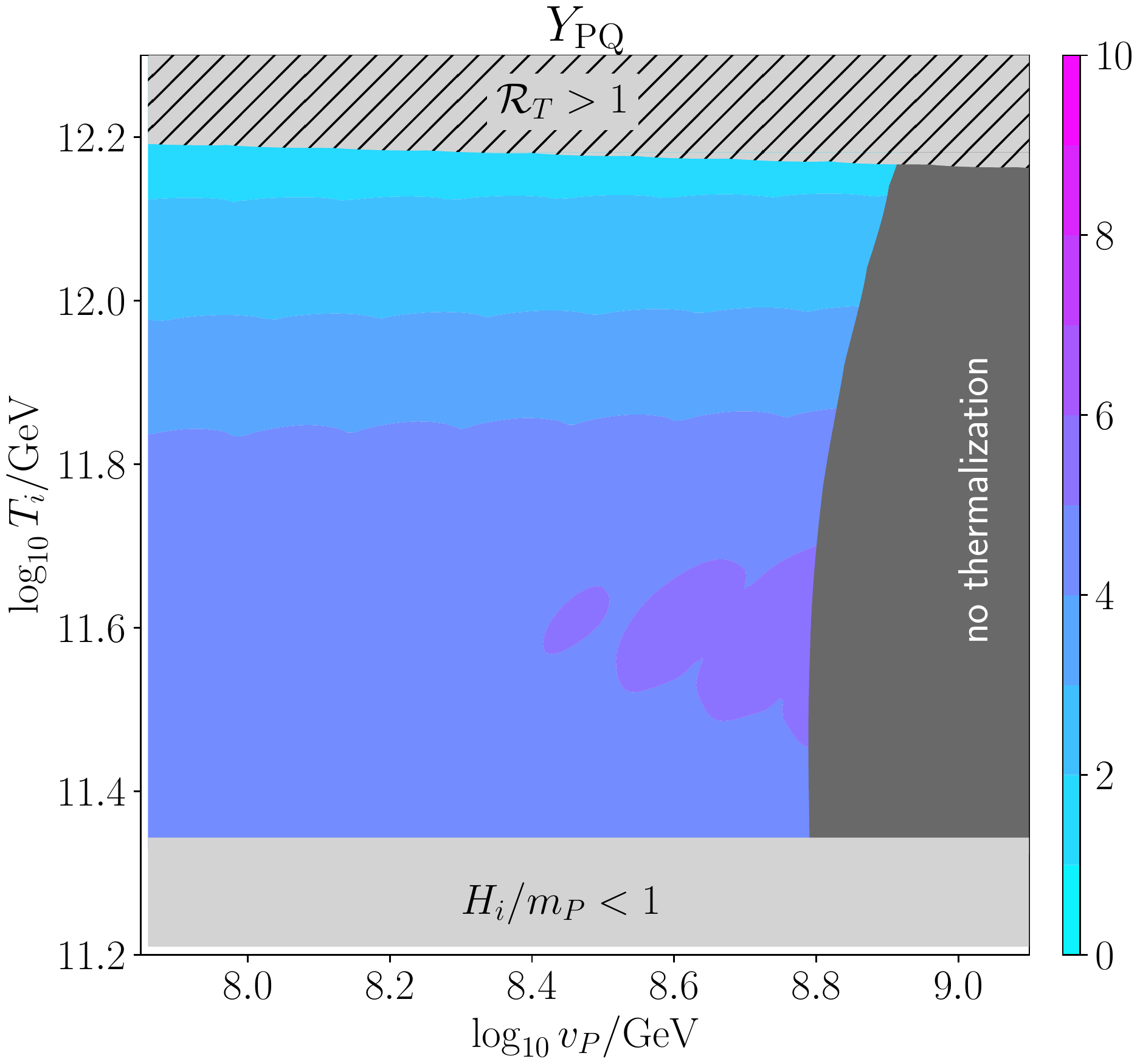}
\end{minipage}
\caption{\label{fig-YPQ}
$Y_\PQ$ after dilution in the $n=10$, $m_P=10^6~\GeV$ ($n=8$, $m_P=10^5~\GeV$)
scenario on the left (right) panel.
The dashed line is $D=10$ 
and the KD era exists in the white line on the left panel.
}
\end{figure}

The scalar potential for the axion $\phi := N_\DW f_a \theta$,
with the axion decay constant, $f_a = \sqrt{2} v_P/N_\DW$, is given by
\begin{align}
 V_a = f_a^2 m_a(T)^2  \left( 1 - \cos \frac{\phi}{f_a}\right),
\end{align}
and the axion mass is approximately given by
\begin{align}
 m_a (T) = 6~\meV \left(\frac{10^{9}~\GeV}{f_a}\right) \times
\begin{cases}
 1 & T \le \LQCD \\
 \left( \dfrac{\LQCD}{T} \right)^p & T \ge \LQCD \\
\end{cases},
\end{align}
where $\LQCD = 150$ MeV is the QCD scale.
Here, we take $p=4$ motivated by the dilute instanton gas approximation~\cite{Gross:1980br}.
We also define the axion mass at zero temperature, $m_{a_0} := m_a(T=0)$.
The axion is produced by the kinetic misalignment mechanism (KMM)~\cite{Co:2019jts}
if the kinetic energy of the axion, $N_\DW^2 f_a^2 \dot{\theta}^2/2$,
is higher than the barrier of the axion potential, $2 f_a^2 m_a(T)^2$,
when the oscillation of the axion starts, $3 H(T_*) = m_a(T_*)$,
where $T_*$ is the temperature at this time.
This condition is satisfied if
\begin{align}
 Y_\PQ = \frac{2\dot{\theta} v_P^2}{s}
 > Y_c :=&\
         N_\DW\times \frac{9}{2} \left(\frac{10}{\pi^2 g_*(T_*)}\right)^{\frac{5}{12}}
           \left(\frac{f_a^{12}}{m_{a_0}\LQCD^4 M_p^7}\right)^{\frac{1}{6}} \\ \notag
        =&\ {0.07} \times {N_\DW}  \times \left(\frac{80}{g_*(T_*)}\right)^{\frac{5}{12}}
                                        \left(\frac{f_a}{10^9~\GeV}\right)^{\frac{13}{6}}.
\end{align}
If this condition is satisfied,
the axion can overcome the potential
and thus the timing to start the oscillation around a minimum is delayed.
Otherwise, the axion will be produced
by the conventional misalignment mechanism~\cite{Preskill:1982cy,Abbott:1982af,Dine:1982ah}.
Thus, the axion density is given by
\begin{align}
&\ \Omega_a h^2 = \frac{s_0 h^2}{\rho_c}  \\ \notag &\ \quad \times
 \begin{cases}
   N_\DW^{-1} C_a  m_{a_0} Y_\PQ
   \sim  0.1\times  N_\DW^{-1} \times
                     \left(\dfrac{10^{8}~\GeV}{f_a}\right)
                     \left(\dfrac{Y_\PQ}{3}\right),
   & Y_\PQ > Y_c, \\
   & \\
   \dfrac{9}{4} \left(\dfrac{10}{\pi^2 g_*}\right)^{\frac{5}{12}}
                \left(\dfrac{m_{a_0}^5 f_a^{12}}{\LQCD^4 M_p^7}  \right)^{\frac{1}{6}}
                \theta^2_*
   \sim 0.11\times \theta_*^2 \left(\dfrac{80}{g_*}\right)^{\frac{5}{12}}
                   \left(\dfrac{f_a}{10^{11.8}~\GeV} \right)^{\frac{7}{6}},
   & Y_\PQ < Y_c,
 \end{cases}
\end{align}
where $\rho_c/s_0 h^2 = 3.6\times 10^{-9}~\GeV$.
Here, $\theta_*$ is the value of $\theta$ when the oscillation starts.
The constant $C_a \simeq 2$ is determined from the numerical calculation
for the delay of oscillation by the kinetic energy~\cite{Co:2019jts}.
The axion density is proportional to the PQ yield, $Y_\PQ$, in the KMM.

The value of $Y_\PQ$ in our model is shown in Fig.~\ref{fig-YPQ}.
In both cases, $Y_\PQ$ is $\order{1}$,
hence $Y_\PQ > Y_c$ is satisfied and the axion density is close to the DM density
when $N_\DW f_a \sim 10^8~\GeV$.
The axion density, in the case of $N_\DW = 1$, is shown in Fig.~\ref{fig-cmn}.
There are regions of parameter space where $\Omega_a > \Omega_\DM$
and the maximum value is about $\xi_a \sim 2$ for $N_\DW = 1$.
However $\Omega_a < \Omega_\DM$ for $N_\DW > 1$ in most of the parameter space.

\subsubsection{LSP density}

\begin{figure}[t]
\centering
\begin{minipage}[t]{0.47\textwidth}
\centering
\includegraphics[height=65mm] {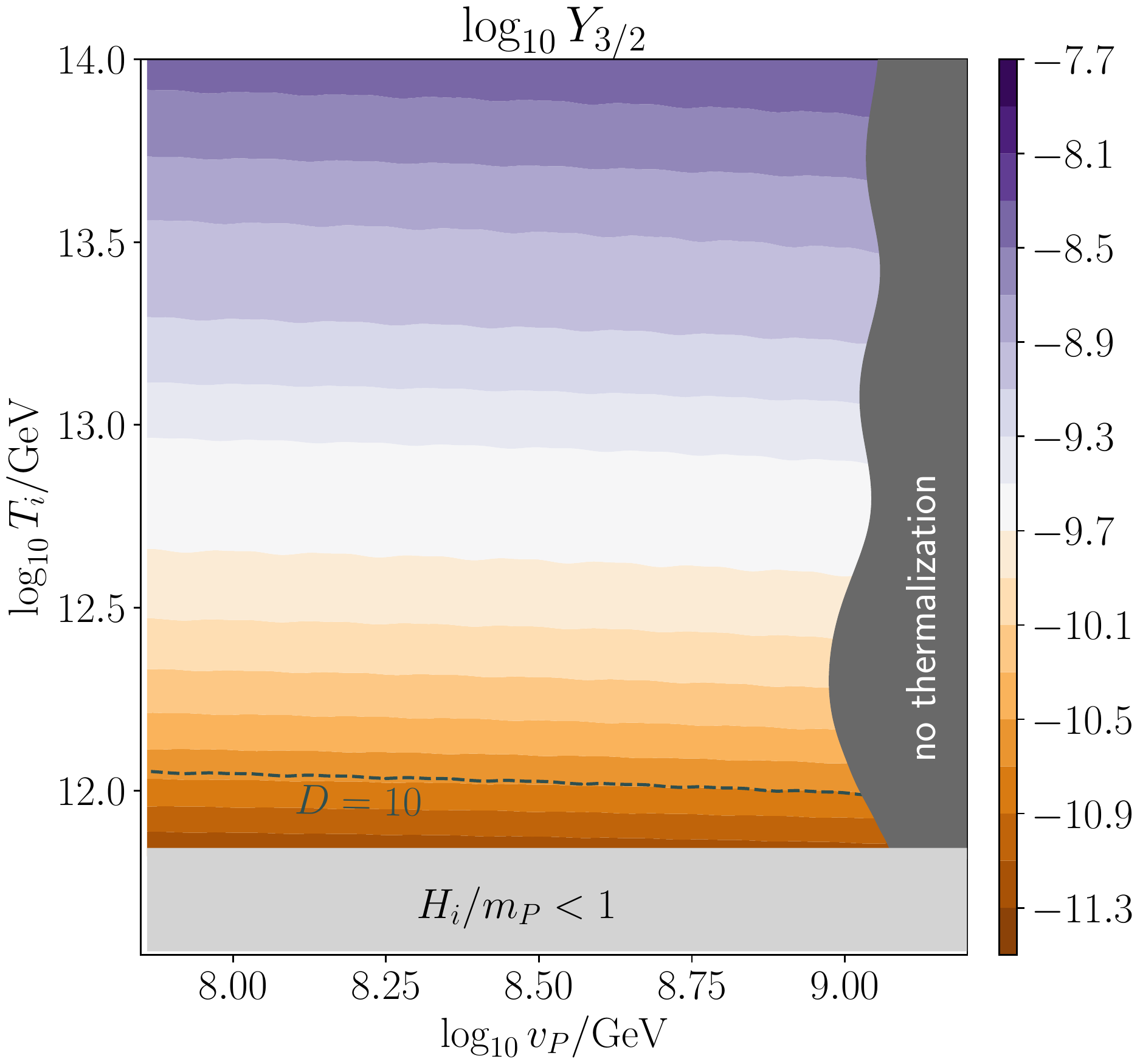}
\end{minipage}
\begin{minipage}[t]{0.49\textwidth}
\centering
\includegraphics[height=65mm] {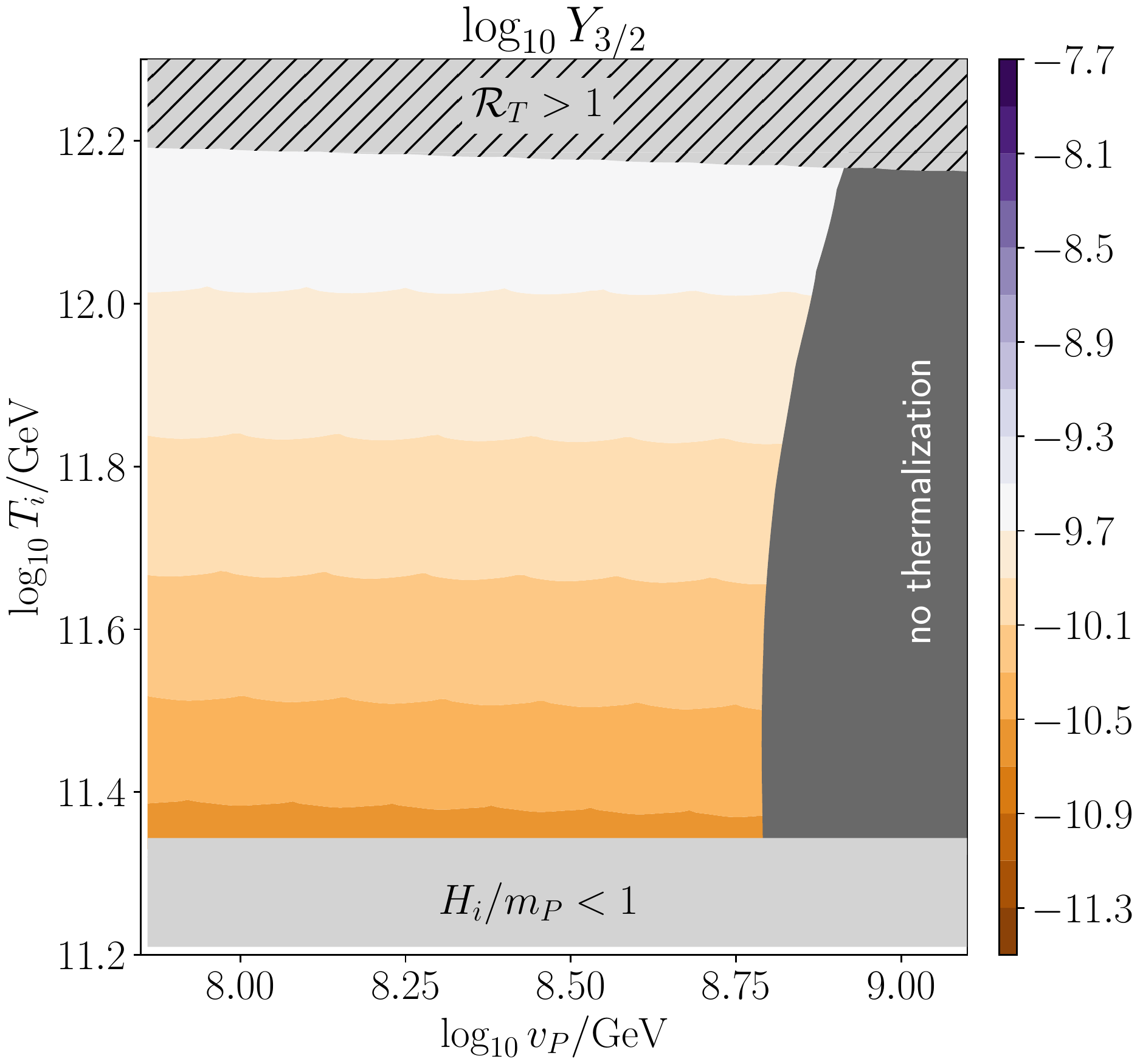}
\end{minipage}
\caption{\label{fig-Y32}
$Y_{3/2}$ after dilution
in the  $n=10$, $m_P=10^6~\GeV$ ($n=8$, $m_P=10^5~\GeV$) scenario
on the left (right) panel.
}
\end{figure}

The neutralino will be the LSP in our scenario
and will be predominantly produced from the late-time decay of the gravitino.
Since we consider a sizable A-term, to produce a sizable PQ asymmetry,
we expect that SUSY breaking is gravity mediated.
Hence, the gravitino mass may be at $\order{m_P}$ or heavier.
In this case, the gravitino is abundantly thermally produced in the early universe
due to the high reheating
temperature $\gtrsim 10^{10}~\GeV$~\cite{Kawasaki:2008qe}.
The Boltzmann equation for the gravitino number density $n_{3/2}$ is given by
\begin{align}
 \dot{n}_{3/2} + 3 H n_{3/2} = C_{3/2},
\end{align}
where the collision term is given by~\cite{Bolz:2000fu,Pradler:2006hh},
\begin{align}
 C_{3/2}(T) = \frac{3 \zeta(3) T^6}{16\pi^3 M_p^2} \sum_{a=1,2,3} c_a g_a^2(T)
\left( 1+\frac{M_a^2(T)}{3m_{3/2}^2}\right) \log \frac{k_a}{g_a(T)}.
\end{align}
Here, $a=1,2,3$ is for the $U(1)_Y$, $SU(2)_L$ and $SU(3)_C$ of the SM.
The constants are $c_a = (11,27,72)$ and $k_a = (1.266, 1.312, 1.271)$.
$g_a$ and $M_a$ are the gauge coupling constants and gaugino masses.
In our numerical analysis, we directly solve this equation assuming $M_a \ll m_{3/2}$.
If the radiation energy dominates the universe when $T=T_i$,
the gravitino is produced at $T=T_i$ and hence
the gravitino yield is approximately given by~\cite{Pradler:2006hh},
\begin{align}
 Y_{3/2} := \frac{n_{3/2}}{s}
\simeq &\ C_{3/2}(T_i) \times D^{-1} \left(\frac{90}{\pi^2 g_*}\right)^{\frac{3}{2}}
                              \frac{M_p}{4T_i^5}    \\ \notag
  \sim&\ 2.0\times 10^{-10} \times  \left(\frac{T_i/D}{10^{12}~\mathrm{GeV}}\right).
\end{align}
If this is not the case,
the gravitino is produced at the end of thermalization.
This is particularly important when the dilution factor is very large.
Figure~\ref{fig-Y32} shows the values of $Y_{3/2}$,
and the gravitino yield is $\order{10^{-14}\mathrm{-}10^{-8}}$ depending
on the reheating temperature and the dilution factor.

The gravitino decay width to MSSM particles is given by~\cite{Moroi:1995fs},
\begin{align}
 \Gamma_{3/2} =  \frac{ 193 m_{3/2}^3}{384 \pi M_p^2},
\end{align}
so the decay temperature of the gravitino
reads
\begin{align}
 T_{3/2}\simeq  &\
\left(\frac{10}{g_*}\right)^{\frac{1}{4}}
      \sqrt{\frac{193m_{3/2}^3}{384\pi^2M_p} }  \\ \notag
           \sim&\
 0.1~\GeV \times \left(\frac{10}{g_*(T_{3/2})}\right)^{\frac{1}{4}}
                 \left(\frac{m_{3/2}}{10^6~\mathrm{GeV}}\right)^{\frac{3}{2}}.
\end{align}
This is typically lower than the temperate at the freeze-out of the neutralino LSP,
$T_f \sim m_\chi/20$~\footnote{
For $m_{3/2} \gtrsim 10^6~\GeV$ assumed in this paper,
the gravitino decays well before the big bang nucleosynthesis~\cite{Weinberg:1982zq,Khlopov:1984pf,Kawasaki:2008qe}.
}.
In our analysis, we assume that the axino mass is heavier than
$\order{m_{3/2}/16\pi^2}$,
and hence the axino decays before the freeze-out of the neutralino~\cite{Choi:2008zq}.
Thus the neutralino LSP produced from axinos does not affect
the LSP density in the current universe~\footnote{
See e.g. Refs~\cite{Rajagopal:1990yx,Choi:2011yf,Kim:2012bb,Choi:2013lwa}
for more discussions about the axino.
}.

The neutralino is copiously produced from gravitino decay
after its thermal freeze-out.  Then it will annihilate if the rate is sufficiently large.
The Boltzmann equation is given by
\begin{align}
\label{eq-Blotzchi}
 \dot{n}_\chi + 3 H n_\chi = - n_\chi^2 \vev{\sigma v}_\chi.
\end{align}
The thermally averaged annihilation rates for the wino and higgsino are
respectively given by~\cite{Olive:1990qm,Olive:1989jg},
\begin{align}
 \vev{\sigma v}_{\wt{W}} \simeq&\ \frac{ 8\pi \alpha_2^2}{M_2^2}
                                  \frac{(1-x_W)^{3/2}}{(2-x_W)^2}, \\
 \vev{\sigma v}_{\wt{H}} \simeq&\ \frac{ 8\pi \alpha_2^2}{\mu^2}
                                  \left(\frac{(1-x_W)^{3/2}}{16(2-x_W)^2}
                                +   \frac{(1-x_Z)^{3/2}}{32c_W^4(2-x_Z)^2}
                                \right),
\end{align}
where $x_V = m_V^2/m_\chi^2$ ($V=W,Z$)
and $c_W = \cos\theta_W$ with the weak angle $\theta_W$.
Here $M_2$ and $\mu$ are the wino and higgsino mass parameters, respectively.
Solving Eq.~\eqref{eq-Blotzchi}, the inverse of the neutralino yield
$Y_\chi := n_\chi/s$,
is given by
\begin{align}
\label{eq-Ychiinv}
 Y_\chi^{-1} \simeq&\
Y_{\chi}(T_{3/2})^{-1}
+ \sqrt{\frac{8\pi^2 g_*(T_{3/2})}{45} } M_p T_{3/2} \vev{\sigma v}_\chi,
\end{align}
where $Y_{\chi}(T_{3/2})$ is the neutralino yield at $T=T_{3/2}$.
$Y_\chi(T_{3/2}) \simeq Y_{3/2}$, assuming that the thermally produced neutralino abundance
is negligible. However,  this is not the case for binos.
The annihilation is effective when the gravitino yield is large, i.e. the
decay temperature is higher and the annihilation rate is larger.

When the annihilation is ineffective, the neutralino density from the gravitino decay
is given by
\begin{align}
 \Omega^{3/2}_\chi h^2 \sim \frac{s_0 h^2}{\rho_c} m_\chi Y_{3/2}
 \sim 0.11 \times
 \left(\frac{m_\chi}{2~\GeV}\right) \left(\frac{T_i/D}{10^{12}~\GeV}\right).
\end{align}
The neutralino should be lighter than $\order{1~\GeV}$
in order to avoid overproduction for $T_i/ D \gtrsim \order{10^{12}~\GeV}$.
This might be the case for the bino LSP,
but the bino LSP will be overproduced thermally in this case
since neither the co-annihilation via nearly degenerate sleptons
nor an enhanced annihilation rate via the mixing
with the wino/higgsino~\cite{Arkani-Hamed:2006wnf} are available.
Thus the annihilation must be effective,
so that the LSP density is not overproduced
for $T_i/ D \gtrsim \order{10^{12}~\GeV}$.
For smaller $T_i/D$,
the LSP production from the gravitino decay become negligible
and the LSP density is governed by the usual thermal freeze-out.

If the annihilation is effective
and the second term in Eq.~\eqref{eq-Ychiinv} dominates,
the DM density is given by
\begin{align}
\Omega_\DM h^2 \sim&\  \xi_\chi^{-1} \Omega_\chi^{3/2} h^2
   \sim\frac{s_0 h^2}{\rho_c}\sqrt{\frac{216}{193}}
                    \left(\frac{10}{g_*}\right)^{\frac{1}{4}}
             \frac{m_\chi \xi_\chi}{\sqrt{m^{3}_{3/2} M_p} \vev{\sigma v}_\chi^\eff}  \\ \notag
             \sim&\ 0.12 \times \left(\frac{10}{g_*}\right)^{\frac{1}{4}}
                               \left(\frac{10^7~\GeV}{m_{3/2}}\right)^{\frac{3}{2}}
                               \left(\frac{m_\chi}{170~\GeV}\right)
                               \left(\frac{\xi_\chi}{0.1}\right)
                               \left(\frac{10^{-26}~\mathrm{cm}^3/s}
                                          {\vev{\sigma v}_\chi^\eff}\right),
\end{align}
where $\vev{\sigma v}_\chi^\eff := \xi^2_\chi \vev{\sigma v}_\chi$
is the effective annihilation cross section
constrained by the indirect detections for the DM.
Thus the DM density can be explained if the gravitino mass is $\order{10^{7}~\GeV}$
which is one or two orders of magnitude larger than $m_P$,
and $\vev{\sigma v}_\chi^{\eff} \sim 10^{-26}~\mathrm{cm}^3/s$
which can be realized by the $\order{100~\GeV}$ wino and higgsino.

\begin{figure}[t]
\centering
\includegraphics[height=70mm] {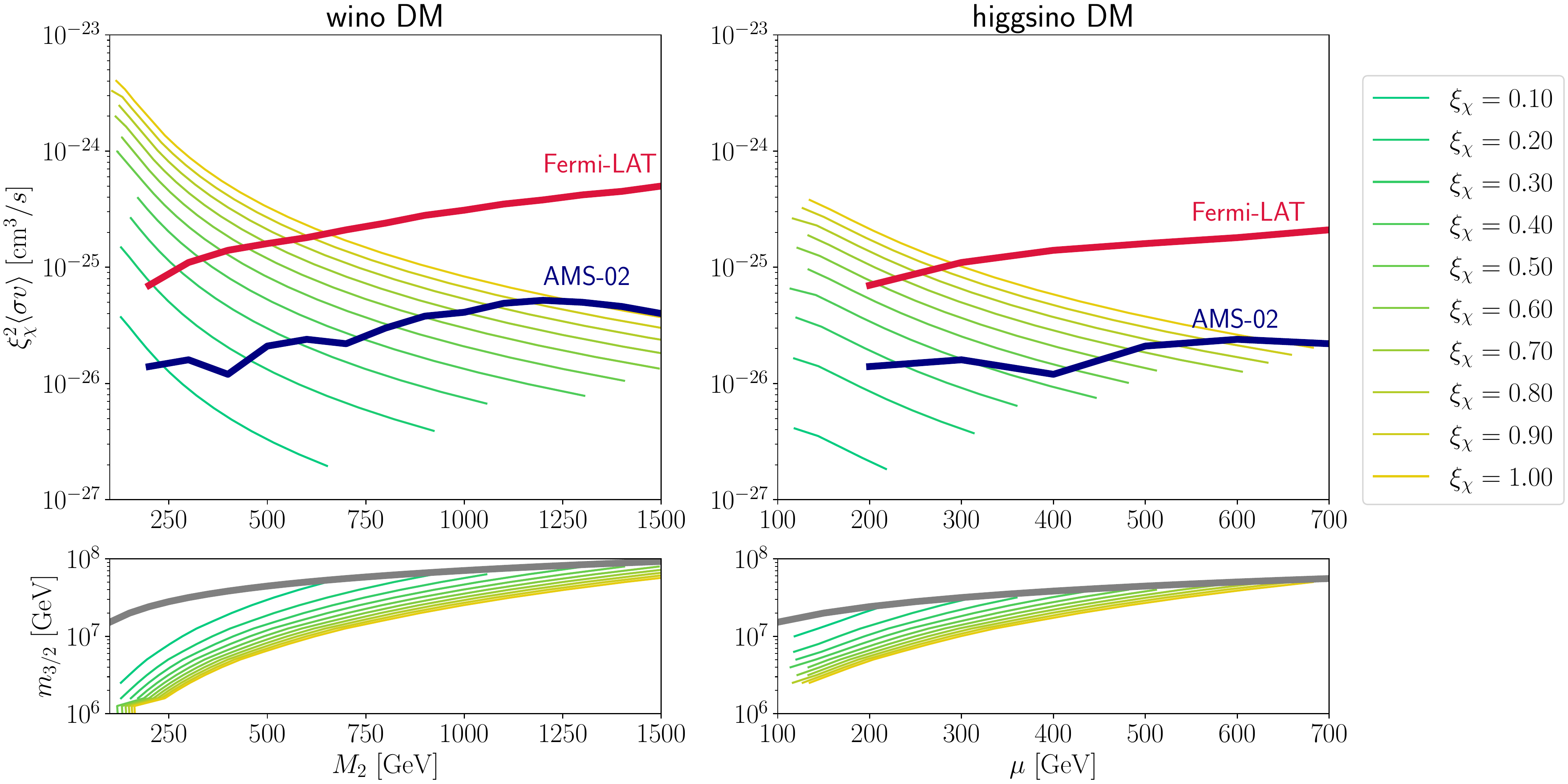}
\caption{\label{fig-LSPindirect}
Indirect detection constraints.
The green lines show the predictions of the LSPs produced from the gravitino decay.
The gravitino mass is chosen such that
$\xi_\chi := \Omega_\chi/\Omega_{\mathrm{DM}}$ is explained.
The red (blue) line is central values of AMS-02 (Fermi-LAT) constraint.
The gravitino masses are shown in the lower panels.
The gray line is $T_{3/2} = T_f = m_\chi/20$,
so the neutralinos are produced by the usual freeze-out mechanism
above this line.
}
\end{figure}

\begin{table}[t]
 \centering
\caption{\label{tab-bench}
Values of various quantities at the benchmark points.
In the last six rows in this table the LSP is assumed to be the wino (the line starting with $M_2$) or the higgsino (the line starting with $\mu$) and the values of the LSP mass, the gravitino mass and the effective cross sections are shown for the correct dark matter abundance.
}
\begin{tabular}[t]{c|cccc} \hline
& point (A) & point (B) & point (C) & point (D) \\  \hline \hline
$n$ & $10$ & $10$ & $8$ & $8$ \\
$N_\mathrm{DW}$ & $1$ & $4$ & $1$ & $2$ \\
$m_P$ [GeV] & $1.0000\times 10^{6}$ & $1.0000\times 10^{6}$ & $1.0000\times 10^{5}$ & $1.0000\times 10^{5}$ \\
$f_a$ [GeV] & $1.4142\times 10^{8}$ & $3.5355\times 10^{8}$ & $1.4142\times 10^{8}$ & $3.9764\times 10^{8}$ \\
$T_i$ [GeV] & $1.0000\times 10^{13}$ & $1.0000\times 10^{13}$ & $1.0000\times 10^{12}$ & $1.0000\times 10^{12}$ \\ \hline  
$c_B m_\nu^2~[\mathrm{eV}^2]$ & $3.2617\times 10^{-4}$ & $6.2633\times 10^{-4}$ & $1.2060\times 10^{-3}$ & $1.5962\times 10^{-3}$ \\
$\Delta N_{\mathrm{eff}}$ & $6.9693\times 10^{-3}$ & $0.0113$ & $3.1958\times 10^{-6}$ & $4.0146\times 10^{-6}$ \\
$D$ & $6.0067$ & $5.1998$ & $1.0665$ & $1.0577$ \\
$Y_{\mathrm{PQ}}$ & $4.7360$ & $5.2712$ & $2.8720$ & $2.9340$ \\
$Y_{3/2}$ & $3.5565\times 10^{-10}$ & $4.1093\times 10^{-10}$ & $1.9290\times 10^{-10}$ & $1.9456\times 10^{-10}$ \\
$\xi_\chi$ & $0.0698$ & $0.8965$ & $0.4359$ & $0.8975$ \\ \hline 
$\mathcal{R}_T$ & $0.0169$ & $0.0190$ & $0.5540$ & $0.5837$ \\ \hline \hline 
$M_2$ [GeV] & $1.9580\times 10^{2}$ & $1.9158\times 10^{3}$ & $1.3423\times 10^{3}$ & $1.9169\times 10^{3}$ \\
$m_{3/2}$ [GeV] & $6.3096\times 10^{6}$ & $1.0000\times 10^{8}$ & $7.9433\times 10^{7}$ & $1.0000\times 10^{8}$ \\
$\langle \sigma v\rangle_{\mathrm{eff}}~[\mathrm{cm}^3/s]$ & $9.5936\times 10^{-27}$ & $1.8290\times 10^{-26}$ & $8.8000\times 10^{-27}$ & $1.8312\times 10^{-26}$ \\ \hline 
$\mu$ [GeV] & $1.1647\times 10^{2}$ & $6.5664\times 10^{2}$ & $4.5887\times 10^{2}$ & $6.5725\times 10^{2}$ \\
$m_{3/2}$ [GeV] & $1.2589\times 10^{7}$ & $5.0119\times 10^{7}$ & $3.9811\times 10^{7}$ & $5.0119\times 10^{7}$ \\
$\langle \sigma v\rangle_{\mathrm{eff}}~[\mathrm{cm}^3/s]$ & $2.0004\times 10^{-27}$ & $1.7652\times 10^{-26}$ & $8.4665\times 10^{-27}$ & $1.7661\times 10^{-26}$ \\ \hline 
\end{tabular}
\end{table}

The indirect detection for DM can probe the annihilation process
originating from DM rich environments, such as Dwarf Spheroidal Galaxies
(dSphs), the Galactic Center and so on~\cite{Leane:2020liq}.
The most relevant limits come
from Fermi-LAT~\cite{Fermi-LAT:2016uux} and AMS-02~\cite{AMS:2016oqu}
which search for gamma ray fluxes from dSphs and anti-proton fluxes, respectively.
Figure~\ref{fig-LSPindirect} shows the effective annihilation cross section
with different values of $\xi_\chi$ in the wino (higgsino) LSP cases
on the left (right) panel.
The gravitino mass is chosen such that the DM density is explained
for a given LSP mass, and its value is shown in the lower panels.
We explore the LSP masses up to $T_{3/2} < T_f \sim m_\chi/20$,
so that the LSP is produced from the gravitino decay.
This limit is shown by the gray line on the lower panels.
For heavier masses $T_{3/2} > T_f$,
the LSP are produced by the usual thermal freeze-out
which has been studied extensively in the literature~\cite{Jungman:1995df,Cirelli:2005uq,Cirelli:2007xd}.
The red (blue) line is the upper bound from the Fermi-LAT (AMS-02) experiment
on the cross section obtained in Ref.~\cite{Cuoco:2017iax}.
Although the central limits from AMS-02 is much stronger,
this could be weaker significantly
due to the propagation uncertainties.
We find that the effective annihilation cross-section can be as low as
$2\times 10^{-27}~\mathrm{cm}^3/s$ for $\xi_\chi \sim 0.1$ and $m_\chi \sim 100~\GeV$.
These light LSPs will be probed by the future experiments
such as the CTA experiment~\cite{CTA:2015yxo,Rinchiuso:2020skh}.

Table~\ref{tab-bench} shows values of various quantities at the benchmark points~\footnote{
A model with $n=10$ and $N_\DW=4$, assumed in the point (B), %s (B) and (C),
is realized in a Pati-Salam unification with
non-anomalous ${Z}_4^R \times{Z}_5$ symmetry~\cite{Kawamura:2020jzb}.
}.
At all the points, the baryon asymmetry is explained with reasonable values of
$c_Bm_\nu^2$ and $\Delta N_\eff$ is smaller than the current limit.
We consider the pure wino and higgsino LSP scenarios,
and the values of the LSP and gravitino masses are shown in the last six rows.
These masses are chosen, so that the DM density is explained
for the given axion densities and
$\vev{\sigma v}_\eff \sim \order{10^{-26}~\mathrm{cm}^3/s}$.
The LSP is predominantly produced from gravitino decay,
so $m_{3/2} \sim \order{10^{7}~\GeV}$
and $m_\chi \sim \order{100\mathrm{-}1000~\GeV}$.
With such large $m_{3/2} \gg m_P$, the A-term $A_P$ may be $\order{m_{3/2}}$.
In this case,
the PQ field may be trapped at a minimum with $\abs{\vev{P}}\gg v_\PQ$,
and thus the dynamics studied in this paper may not happen,
see e.g. Refs.~\cite{Kawasaki:2000ye,Kawasaki:2006yb}.
Therefore, the relation assumed in this paper, $A_P \simeq m_P$,
should hold even if $m_{3/2} \gg A_P$.
This is a requirement for a SUSY breaking mediation scenario.

The wino and higgsino above 100 GeV are not excluded by the LEP experiment~\cite{ALEPH:2002gap},
but these can be tested by the LHC and future colliders.
If the LSP is purely wino or higgsino,
the searches for disappearing tracks are available~\cite{Ibe:2006de,Mahbubani:2017gjh,Fukuda:2017jmk}
and the current limits are 660 (210) GeV
for the wino (higgsino) LSP mass~\cite{ATLAS-CONF-2021-015}.
Hence, the benchmark (A) in Table~\ref{tab-bench}
is excluded for both cases of wino and higgsino LSP.
The limits are relaxed if the LSP is a mixture
and the lifetime of the chargino is shorter.
This can be more easily achieved in the higgsino LSP case,
when the mass difference between the chargino and the LSP is $\order{1~\GeV}$.
The $\order{1~\GeV}$ mass difference can be achieved
if e.g. the wino mass is at sub-TeV~\cite{Kawamura:2017amp}.
For the mixed LSP case~\cite{Arkani-Hamed:2006wnf}, however,
the direct searches for DM give stronger constraints,
and hence this case would be tested by near future observations.
The detailed study about the constraints from the LHC and direct searches
are beyond the scope of this paper.

\section{Summary}
\label{sec-Summary}

In this paper,
we studied the lepto-axiogenesis scenario in the minimal SUSY KSVZ axion model
with the type-I see-saw mechanism.
We developed a way to follow the PQ field dynamics from the beginning of the rotation
to the approach to the minimum.
While the rotation is not too fast,
we can directly follow the dynamics of the PQ field by solving Eq.~\eqref{eq-deqys}.
The evaluation becomes, however,
less efficient for later times due to the extremely fast rotation,
hence we trace the dynamics by averaging over the rotation
based on the ansatz Eq.~\eqref{eq-solpsi},
and the solution is given by Eq.~\eqref{eq-semianal}.
We find the thermalization is well described by $\Delta$
which represents the ellipticity of the rotational motion.
The evolution of $\Delta$, together with the radiation energies,
can be calculated by solving
Eqs.~\eqref{eq-deaDp}, \eqref{eq-deaRp} and \eqref{eq-deaAp}.
Based on the solution obtained by directly solving the equations of motion,
we can evaluate the averaged values of the amplitude of the PQ field,
energy densities, angular velocity and so on.
The solutions and evolution equations solved numerically
do not rely on the form of the dominant energy density of the universe.
Thus our solution is applicable for the case
when the radiation and the PQ field energy are comparable,
such as the case shown in the left panel of Fig.~\ref{fig-evolrho}.
Although we focus on the minimal KSVZ model for illustration, a
similar analysis could be applied for models with more PQ fields and/or
different thermalization mechanisms which appear in e.g. DFSZ model~\cite{Zhitnitsky:1980tq,Dine:1981rt}.

We studied the baryon asymmetry, $\Delta N_\eff$ and the DM density based
on the PQ field dynamics.
When $n=10$, the matter domination era always occurs
and the baryon asymmetry is predominantly produced at the end of thermalization.
The soft mass for the PQ field, $m_P$, should be $\order{10^6~\GeV}$ in this case in order
to explain the correct amount of the baryon asymmetry.
Although there is matter domination, the produced PQ asymmetry or gravitino yield
are not diluted significantly since the dilution factor is maximally of $\order{10}$
when $T_i \gtrsim 10^{12}~\GeV$ and $H_i > m_P$.
Thus the gravitino tends to be produced abundantly
and the LSP annihilation must be effective in order to avoid overproduction of the LSP.
The DM density is explained if the gravitino mass is $\order{10^7~\GeV}$
and the wino (higgsino) mass is $\order{10^3~\GeV}$ ($\order{100~\GeV}$).
The light neutralino DM will be tested by future indirect detection experiments,
such as CTA.
Since there is a region of parameter space where $\Delta N_\eff \sim \order{0.01}$,
future experiments would be able to probe this scenario.
In addition, kination energy can be a dominant or sizable
component of the total energy,
and hence this could be seen in the gravitational wave spectrum~\cite{Co:2021lkc,Gouttenoire:2021wzu}.

When $n=8$,
the matter domination epoch is absent for a sufficiently large initial temperature,
and hence $m_P \sim \order{10^5~\GeV}$ can explain the baryon asymmetry.
The favored mass range for the gravitino and neutralino are similar to the case of $n=10$.

\section*{Acknowledgment}
The work of J.K.
is supported in part by
the Institute for Basic Science (IBS-R018-D1),
and the Grant-in-Aid for Scientific Research from the
Ministry of Education, Science, Sports and Culture (MEXT), Japan No.\ 18K13534.
The work of S.R.is supported in part by the Department of Energy (DOE) under Award No.\ DE-SC0011726.

\appendix
\section{Note for $H_i < m_P$}
\label{sec-HilemP}

In the main text, we focus on the case with $H_i > m_P$
and the kick by the A-term occurs after the reheating ends.
We shall briefly discuss the case of $H_i < m_P$ in this Appendix.
We need to specify a model of inflation and reheating
to study the PQ field dynamics in the same manner as in the main text,
so this is beyond the scope of this paper.

When $H_i < m_P$,
the PQ starts to rotate during matter domination by an inflaton
before the reheating ends.
For simplicity, let us assume the reheating occurs instantly
and the Hubble parameter is given by
\begin{align}
 H = H_i e^{-3u/2},
\end{align}
where $u < 0$. Here we take $u=0$ at the time of reheating.
When $H > m_P$ ($> H_i$),
\begin{align}
 \abs{P} \sim \left(\frac{c_H H^2 M_p^{2n-6}}{(n-1)\la^2}\right)^\frac{1}{2n-4}
         = \abs{P_\osc} \left(\frac{H}{m_P}\right)^{\frac{1}{n-2}},
\end{align}
where
\begin{align}
 \abs{P_\osc} := \left(\frac{c_H m_P^2 M_p^{2n-6}}{(n-1)\la^2}\right)^\frac{1}{2n-4}.
\end{align}
Then,
the PQ field starts to rotate at $H \sim m_P$
and the amplitude scales as
\begin{align}
 \abs{P} = \abs{P_\osc} \frac{H}{m_P}
\end{align}
when $H < m_P$.
Therefore the amplitude at $u=0$ is given by
\begin{align}
\abs{P_i}\sim \frac{H_i}{m_P} \abs{P_\osc} =
   \frac{H_i}{m_P}\left(\frac{c_H m_P^2 M_p^{2n-6}}{\la^2}\right)^\frac{1}{2n-4}.
\end{align}

The PQ asymmetry is dominantly generated when the rotation starts
as in the case of radiation domination.
Using the above scaling laws, the PQ number at $u=0$ is estimated as
\begin{align}
 n_\PQ(u=0) 
           \sim \frac{2n^2}{3n-9}
            \left(\frac{c_H}{(n-1)\la^2}\right)^{\frac{n}{2n-4}}
            \left(\frac{M_p}{m_P} \right)^{\frac{2n-6}{n-2}}
            A_P H_i^2 \sin n\theta_\osc,
\end{align}
hence the PQ yield is given by
\begin{align}
\label{eq-YPQinf}
 Y_\PQ(0) \sim&\
            \frac{n^2}{6(n-3)} \left(\frac{c_H}{(n-1)\la^2}\right)^{\frac{n}{2n-4}}
            \left(\frac{M_p}{m_P} \right)^{\frac{2n-6}{n-2}}
            \frac{A_P T_i}{M_p^2} \sin n\theta_\osc \\ \notag
    \sim&\ 48 \times
            \left(\frac{c_H}{9\la^2}\right)^{\frac{5}{8}}
            \left(\frac{1~\PeV}{m_P} \right)^{\frac{7}{4}}
            \left(\frac{A_P}{1~\PeV}\right)
            \left(\frac{T_i}{10^{11}~\GeV}\right)
           \sin n\theta_\osc.
\end{align}
We assume $n = 10$ in the second equality.
Thus, the PQ yield is proportional to $T_i$ if $H_i < m_P$.

The baryon asymmetry produced by the rotation of the PQ field
will be evaluated in the same manner as in the case of $H_i > m_P$,
since the asymmetry produced before the reheating is diluted away.
During the (first) radiation dominant era,
the baryon asymmetry is evaluated as in Eq.~\eqref{eq-YBRD}
with $w_L=w(u_L)$ replaced by $w(0)$ if $T_i < T_L$.
If there is matter domination by the PQ field,
the asymmetry is evaluated as Eq.~\eqref{eq-IntMD}.
Neglecting the mild dependence through $w$ and $\Delta$,
both of Eqs.~\eqref{eq-YBRD} and~\eqref{eq-IntMD}
are independent of $T_i$ and $Y_\PQ$.
Thus the baryon asymmetry will not be changed significantly
from the cases with $H_i > m_P$
as long as the kick effect is so large such that $w(0) \gtrsim 1$ after reheating.
The baryon asymmetry would be too small
if the kick effect is not sufficient such that $w(0) \ll 1$.
Lower values of $T_i$ may be favored in order to explain the baryon asymmetry and the DM density
simultaneously, since the gravitino yield is suppressed by $T_i$
but the baryon asymmetry is sizable.
Studying lepto-axiogenesis with a concrete inflation scenario
with a low reheat temperature is our future work.

\section{Thermal log potential}
\label{sec-ThLog}
\newcommand{\tm}{\widetilde{m}_P}

\begin{figure}[t]
\centering
\begin{minipage}[t]{0.48\textwidth}
\centering
\includegraphics[height=63mm] {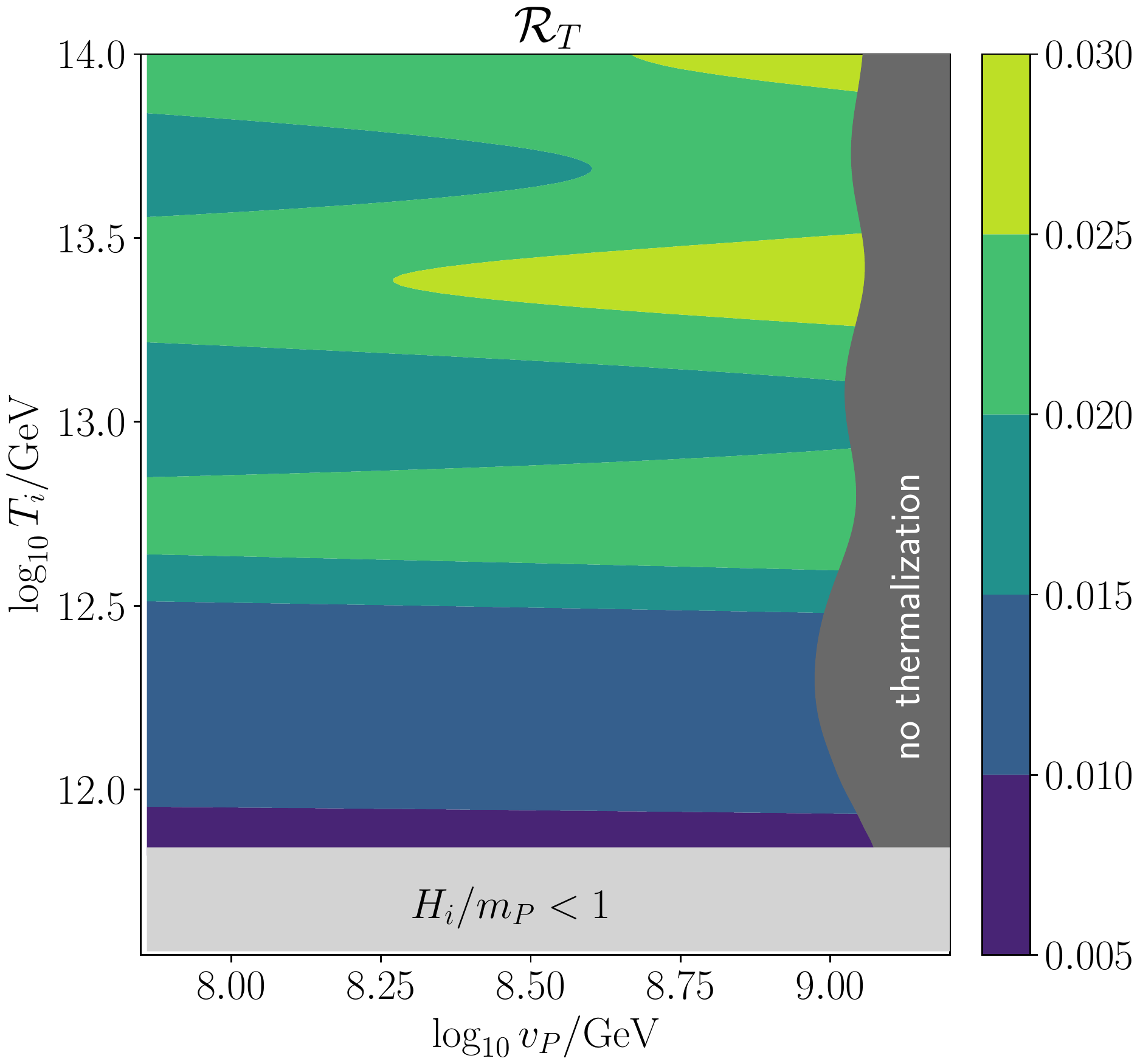}
\end{minipage}
\begin{minipage}[t]{0.48\textwidth}
\centering
\includegraphics[height=63mm] {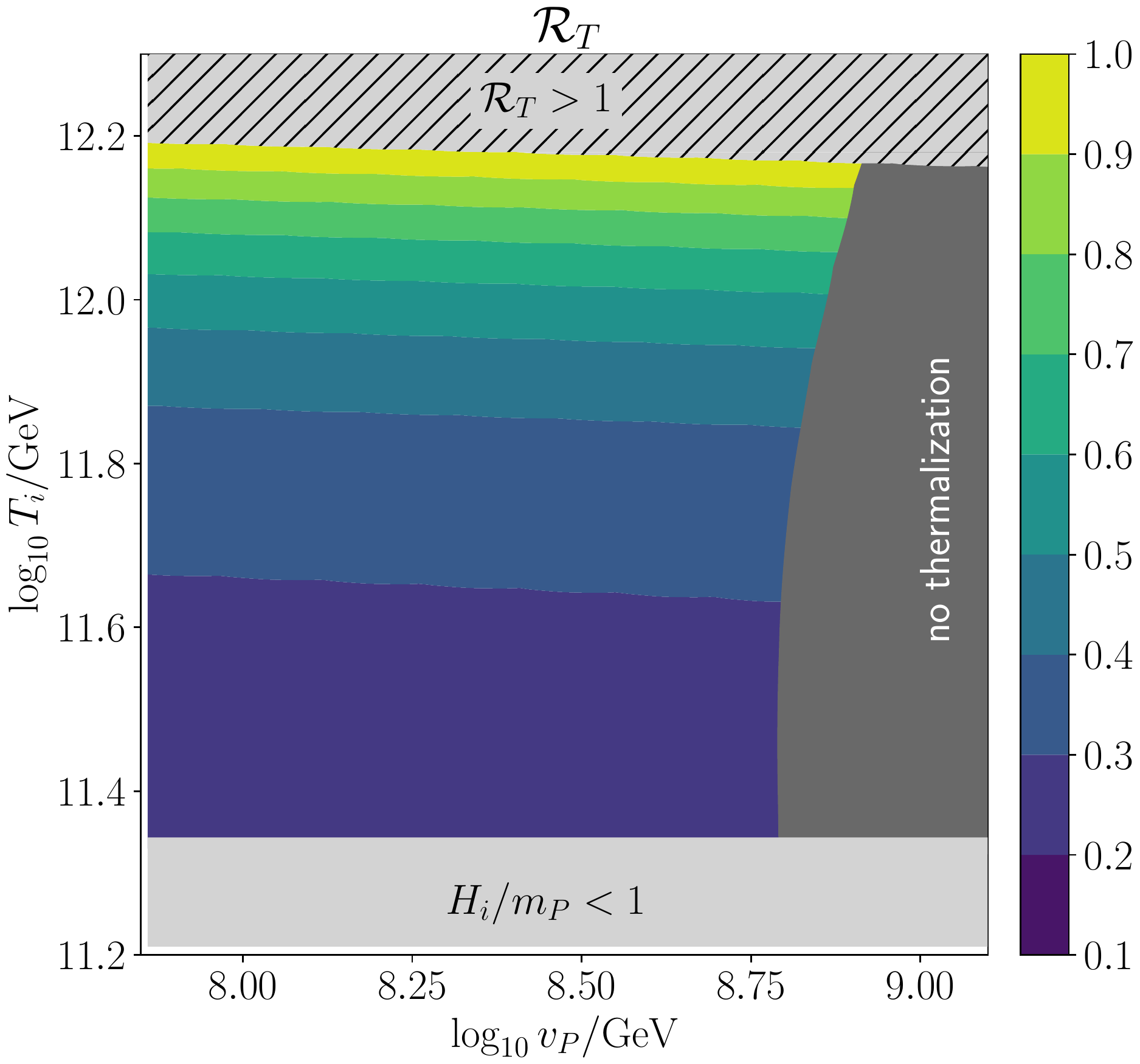}
\end{minipage}
\caption{\label{fig-RT}
$\mathcal{R}_{T}$ in the  $n=10$ ($n=8$) scenario on the left (right) panel.
}
\end{figure}

Let us discuss how the thermal log potential affects the PQ field dynamics.
Before the rotation starts,
the thermal log term is negligible compared
with the self-coupling $\propto \abs{P}^{2n-2}$ and the A-term
due to the large value of $\abs{P_i}$.
Then,
the thermal-log potential may be relevant
after the rotation around the minimum starts.
With only the mass term, the derivative of the potential is given by
\begin{align}
 \frac{\partial V}{\partial P^*} =
\left(\tm^2 + a_L \alpha_s^2 \frac{T^4}{\abs{P}^2}\right) P,
\quad
\tm^2 := m_P^2 \log\frac{\abs{P}^2}{v_P^2}
\end{align}
We define the relative importance of the thermal-log term, $R_T$, as
\begin{align}
 R_T := a_L \alpha_s^2 \frac{T^4}{\tm^2 \abs{P}^2}.  \label{eq:RT}
\end{align}
Assuming radiation domination, the value of $R_T$ when the rotation starts at $H = \tm$, is given by
\begin{align}
 R_T(u_\osc) \sim \frac{90a_L \alpha_s^2}{\pi^2 g_*}
 \left(\frac{(n-1)\la^2 M_p^2}{c_H \tm^2}\right)^{\frac{1}{n-2}}.
\end{align}
When $(n,m_P) = (10,~10^{6}~\GeV)$ and $(8,~10^{5}~\GeV)$,
$R_T(u_\osc) = 0.65$ and $15.9$, respectively. 
Here, $\alpha_s = 0.1$ is assumed. 
Thus, the thermal-log term will not be negligible at $u \sim u_\osc$.
After the rotation starts, $R_T$ scales as $e^{-u}$
and the thermal-log term becomes less important at later times.

For $u>u_1$, we shall consider the relative importance of the thermal-log effects
for the averaged values,  
i.e.
\begin{align}
R_T
= \zeta(T) \times  \frac{e^{-2\Omega}}{(4\Omega+2\Delta) \sinh\Delta}
\simeq \zeta(T) \times \frac{2 e^{3(u-u_1)}}{\sqrt{w} C_w},
\end{align}
where the analytical solution Eq.~\eqref{eq-Asol}
is used in the second equality.
Figure~\ref{fig-RT} shows the maximum value of $R_T$ in $u \in [u_1, u_K]$,
\begin{align}
\label{eq-defRT}
 \mathcal{R}_T := \max_{u_1 < u < u_K} R_T(u).
\end{align}
We evaluated the maximum value,
but we found $\mathcal{R}_T = R_T(u_1)$ in the parameter space
because of the short MD era.
We see that $\mathcal{R}_T \lesssim 0.03$ in the case of $n=10$, $m_P=10^6~\GeV$
and thus the thermal-log effect is negligible. 
It can be, however, sizable for $n=8$.
For $T_i \lesssim 10^{12}~\GeV$,
$\mathcal{R}_T \lesssim 0.5$ and the thermal-log effect is sub-dominant,
while for higher $T_i$,
$\mathcal{R}_T \gtrsim 0.5$ and it can be larger than unity in a wide region of parameter space.
Thus we need to take the thermal-log effect into account
for $n=8$ and $T_i \gtrsim 10^{12}~\GeV$ in order to improve the accuracy.
The values of $\mathcal{R}_T$ at the benchmark points are shown in Table.~\ref{tab-bench}.

{\small
\bibliographystyle{JHEP}
\bibliography{PSleptogenesis}
}
%%%%%%%%%%%%%%%%%%%%%%%%%%%%%%%%%%%%%%%%

\end{document}